\RequirePackage{fix-cm}
\documentclass[smallextended]{svjour3}       
\smartqed  
\usepackage{graphicx}
\usepackage[utf8]{inputenc}
\usepackage{amsmath}

\usepackage{amsthm}
\usepackage{amssymb}
\usepackage{verbatim}
\usepackage{tikz-cd}
\usepackage{caption}
\usepackage[mathscr]{euscript}
\usepackage{csquotes}
\usepackage[doi=false, url=false, isbn=false, eprint=false, backend=biber,style=authoryear,maxcitenames=6, maxbibnames=99]{biblatex}
\newcommand{\ra}{\rightarrow}

\journalname{Social Choice and Welfare}
\begin{document}

\title{Proxy Selection in Transitive Proxy Voting}


\author{Jacqueline Harding}

\institute{Jacqueline Harding \at
              Stanford University \\
              \email{hardingj@stanford.edu}
}

\date{}

\maketitle

\begin{abstract}
Transitive proxy voting (or `liquid democracy') is a novel form of collective decision making, often framed as an attractive hybrid of direct and representative democracy. Although the ideas behind liquid democracy have garnered widespread support, there have been relatively few attempts to model it formally.

This paper makes three main contributions. First, it proposes a new social choice-theoretic model of liquid democracy, which is distinguished by taking a richer formal perspective on the process by which a voter chooses a proxy. Second, it examines the model from an axiomatic perspective, proving (a) a proxy vote analogue of May's Theorem and (b) an impossibility result concerning monotonicity properties in a proxy vote setting. Third, it explores the topic of manipulation in transitive proxy votes. Two forms of manipulation specific to the proxy vote setting are defined, and it is shown that manipulation occurs in strictly more cases in proxy votes than in classical votes.
\end{abstract}

\section{Introduction}
\label{introduction}
Transitive proxy voting (or `liquid democracy') is a novel form of collective decision making. It is often framed as an attractive hybrid of direct and representative democracy, purporting to balance pragmatic factors with the ability to represent a population. Recently, it has been used by the German branch of the Pirate Party to aid intra-party decisions \parencite{Litvinenko2012}. Although the ideas behind liquid democracy have garnered widespread support, there has been little rigorous examination of the arguments offered on its behalf. In particular, there have been relatively few attempts to model liquid democracy formally (discussed in Section \ref{related work}). A formal model has the potential to serve as a testing ground for the conceptual and empirical claims put forward by supporters of liquid democracy.

This paper attempts to fill this gap, presenting a new model of liquid democracy. The model is distinguished by the fact that it takes a richer formal perspective on proxy selection, the process by which a voter chooses a proxy; it is argued that this allows it better to capture features relevant to claims made about liquid democracy. The model is examined from an axiomatic perspective, then put to work exploring the hitherto undeveloped topic of manipulation in a proxy vote setting.

\citeauthor{Blum2016} (\citeyear{Blum2016}, p.165) characterise liquid democracy as the conjunction of four principles. A voter can (a) vote directly, (b) delegate her vote to a representative to vote on her behalf (this representative is called her `proxy'), (c)
delegate those votes she has received via delegation to another representative and; (d) terminate the delegation of her vote at any time. It is argued that the flexibility of liquid democracy confers it certain benefits which classical voting --- where voters vote directly --- lacks. Let us recall some of these benefits here.

\paragraph{Balancing Practicality and Democracy.} Direct democracy, where citizens vote directly on issues through frequent referenda, is seen as `strongly democratic but highly impractical' (\cite{Green-Armytage2015}, p.190), whilst representative democracy, where citizens elect representatives to make decisions on their behalf, is `practical but democratic to a lesser degree' (\cite{Green-Armytage2015}, p.190). Under proxy voting, no such trade-off need occur. If people want their particular views to be represented in a vote, they can ensure this by voting directly. If they are undecided on an issue (or practical factors prevent them from becoming sufficiently informed, or even from casting their vote directly), they can choose to delegate their vote to someone they perceive as competent or trustworthy.

\paragraph{Increasing Voter Turnout.} There are at least three arguments for the claim that proxy voting increases voter turnout. Firstly, \textcite{Miller1969} argues that a major barrier to voters' participation in elections is simply the opportunity cost of voting directly. Secondly, both \textcite{Miller1969} and \textcite{Alger2006} identify apathy towards political representatives as a reason for poor voter turnout. If a voter can be represented by someone whom she trusts, they argue, she will be more likely to vote. Since proxy voting (at least as normally construed) allows voters to delegate their votes to any other voter, it seems more likely that voters will be represented by someone they approve of. Thirdly, voters are often deterred from voting by the fact that they haven't made their mind up about all the alternatives being considered in the election (even if they have some sense of what they think). By choosing their proxy carefully, they can vote on some alternatives but not others. \textcite{Behrens2017} argues that the `transitivity' of liquid democracy (where proxies can delegate the votes they have been given) accentuates this benefit, since it can only increase the number of potential representatives for a voter.

\paragraph{Increasing Competence of Voters.} A voter might delegate her vote when she believes another voter to be better informed than her. Assuming this perception of competence is truth-tracking, \textcite{Green-Armytage2015} argues that this implies that proxy voting leads to votes being cast by voters who are (on average) better informed.

\paragraph{Increasing Diversity of Viewpoints.} \textcite{Alger2006} observes that proxy voting might lead to greater diversity in the viewpoints expressed by voters. In a representative democracy, only a very small proportion of the population is a potential representative; this means that such representatives tend to be pushed towards viewpoints with more broad appeal. By increasing the number of potential representatives, proxy voting could allow voters to express more idiosyncratic viewpoints.
\\
\\
\noindent The ability to represent the benefits outlined above is a desideratum of a formal model of liquid democracy. In Section \ref{related work}, existing models of liquid democracy from within the social choice literature are surveyed. It is argued that none of these models properly represents the means by which a voter chooses a proxy. This motivates the model of transitive proxy voting proposed in Section \ref{introducing the model}, which is distinguished by its ability to include preference information in proxy selection. Formally, the model presented augments a classical vote $(N,A,f)$ (where $N$ is a set of voters, $A$ is a set of alternatives and $f$ is a resolute social choice function) with a novel function $g$, called a `proxy mechanism'. In Section \ref{properties of proxy mechanisms}, novel properties of proxy mechanisms $g$ are explored, and a natural proxy mechanism (the \texttt{SUBSET} mechanism) is characterised using some of these properties.  In Section \ref{properties of proxy votes}, properties of pairs $(f,g)$ are defined. A proxy vote analogue of May's Theorem (which characterises the majority rule when $|A|=2$) is proved and an impossibility result in a proxy vote setting is presented and proved, showing that certain desirable properties of pairs $(f,g)$ are incompatible with natural properties of their individual components $f$ and $g$. In Section \ref{manipulation}, manipulation in a proxy vote setting is examined. A novel form of manipulation (`proxy choice manipulability') is defined and connections between this form of manipulation and classical manipulation are explored. The effect of single peakedness on manipulation in a proxy vote setting is also examined; it is demonstrated that strategyproofness is strictly harder to come by in proxy elections.

\section{Related Work}
\label{related work}
This section outlines existing social-choice theoretic models of transitive proxy voting. Discussion of these models motivates a model which takes a more robust perspective on proxy selection; such a model is introduced in the next section.

\subsection{Pairwise Delegation}

\textcite{Brill2018PairwiseLD} observe that we can view ordinal preference rankings as collections of pairwise comparisons (or `edges') between alternatives. When voters provide linear orders $\succ$ over some set of alternatives $A$, they are effectively choosing whether $a \succ b$ or $b \succ a$ for each $a,b\in A$. In the ordinal preference setting, then, this allows us to model voters with partial opinions as having fixed some edges but not others. Similarly, \textcite{Christoff17} have voters vote or delegate on interdependent binary issues. This model generalises that of \textcite{Brill2018PairwiseLD}; once we translate ordinal rankings into a binary aggregation setting, pairwise comparisons become binary issues.

In both \citeauthor{Brill2018PairwiseLD}'s and \citeauthor{Christoff17}'s models, for each pair of alternatives $(a,b)$ (or each binary issue $p$) that a voter has not decided between, she chooses some delegate from amongst the other voters to decide on her behalf whether $a \succ b$ or $b\succ a$. So her delegations are `pairwise'; a voter might have a different delegate for each edge she is undecided on, meaning she ends up submitting an intransitive preference order (or, in the more general aggregation setting, a ballot violating some rationality requirement). This means that we must either modify the social choice function to accommodate intransitivity or provide a systematic way of forming preference profiles from the outputs of delegations.

Of course, the issue with intransitivity arises only when we allow pairwise delegations. The model I propose in this paper will restrict delegations such that each voter picks at most one proxy. This loss of generality is motivated not merely from a desire to circumvent issues of intransitivity but also by consideration of proxy selection, the process by which a voter chooses a proxy. For example, according to \citeauthor{Brill2018PairwiseLD}, delegation is done on the basis of the perception of competence. But if we accept that it is irrational to hold intransitive preferences oneself (an assumption which I won't challenge here), then it is unclear why we should think it rational to accept an intransitive preference resulting from delegation. Surely the conclusion a voter ought to draw when her delegates present her with an intransitive ballot is that she was mistaken in her initial assessment of the competence of her delegates? It appears that --- by allowing pairwise delegations in the absence of a well-developed account of proxy selection --- we are condoning irrationality at a distance.

\subsection{Proxies Represent Voters' Interests}

In the model proposed by \textcite{Bloembergen2018OnRD}, voters in a network choose between two alternatives. For each voter, one alternative is better, but voters are not aware which alternative is better for any of the voters, including themselves. If a voter votes directly, there is a publicly-known probability that she will vote for the alternative which is worse for her. Each voter either votes directly or delegates her vote to one of her neighbours in the network (so delegations are transitive). Note that since voters are aware of neither their interests nor other voters', it's possible for a voter to delegate to a neighbour whose interests aren't aligned with hers. The utility a voter gains from voting is proportional to the probability that the voter who ends up casting her vote --- her terminal delegate, or `guru' --- votes for the alternative which is better for her.

In the model proposed by \textcite{Abramowitz2018}, an electorate votes on a set of binary issues. The electorate is composed of voters and delegates; voter preferences over the issues are private, but delegate preferences are public. Voters then express preferences over delegates; a voter's attitude towards a delegate is assumed to correspond solely with the degree of agreement between the voter's and delegate's preferences over the issues.

In \citeauthor{Abramowitz2018}'s model, proxy selection depends on a perception of correspondence between a voter's views and those of her proxy (this could also be said of \citeauthor{Bloembergen2018OnRD}'s model; in their case, a perception of correspondence can be mistaken); this seems to me an essential component of any account of proxy selection, and my model will incorporate it. The models lack, though, a representation of all of the other factors that can make a voter choose a proxy (knowledge, charisma, etc); we saw above that these factors are important in motivating transitive proxy voting.

\subsection{Preferences over Delegates}

In the model proposed by \textcite{escoffier_convergence_iterative}, each voter $i \in N$ (where $N$ is the set of voters) submits a preference ordering $\succ_{i}$ over $N\cup \{0\}$, interpreted as representing who $i$ would prefer to end up casting her vote (i.e. a preference relation over her potential gurus), with `$0$' representing the possibility of abstention.

Similarly, \textcite{Kotsialou2018} propose a model where voters submit preferences over the set of alternatives $A$ or preferences over $N$ (the latter interpreted as a preference over their immediate proxy, rather than their guru). In the first case, voters are taken to cast their own vote. In the second case, voters are taken to delegate their vote, with the delegate decided on by a central mechanism.

By having voters rank other voters, both models neatly represent the idea that several factors can inform a voter's choice of proxy. What is missing from the models, though, is an important component of proxy selection we discussed above, namely that a voter's choice of proxy ought to depend on correspondence between her views and the proxy's. In \citeauthor{escoffier_convergence_iterative}'s model, for example, there is no actual election in which voters are participating, meaning there is no way to model such correspondence. Similarly, \citeauthor{Kotsialou2018}'s model takes an all-or-nothing approach to delegation. Voters are immediately categorised into direct voters or delegators, regardless of the actual content of the preferences they submit (the existence of a preference order of either sort is sufficient to determine this categorisation). So the model is unable to allow voters to express preferences on some issues but not others, and relate proxy selection to these partial preferences.

\subsection{Ground Truth}

The models I've discussed above take a voting procedure to aggregate the preferences of an electorate; the model I propose in this paper is of this sort. A very different sort of model measures the accuracy of a voting procedure relative to an underlying ground truth. For example, \textcite{Cohensius2017} consider an infinite population $N$ of voters distributed on a real interval $[a,b]$; a distinguished finite $N'\subseteq N$ are allowed to cast their votes directly (a vote consists of reporting their position on the interval), whilst each other voter delegates her vote to the closest voter in $N'$ (so delegation is non-transitive). The authors find that proxy voting is always more accurate than direct voting when the ground truth is taken to be the taken to be the median of the voters' positions, and generally (through simulation) more accurate when the ground truth is the mean of the voters' positions.

Similarly, \textcite{Kang2018} consider a vote on a binary issue, for which it is assumed there is a ground truth. $N$ voters are arranged in a social network. Similarly to \textcite{Bloembergen2018OnRD}, each voter $i \in N$ has a public competence level $p_{i}$, interpreted as the probability she would vote correctly if she voted directly. Voters can either vote directly or delegate their vote to a neighbour in the network whose competence level is strictly higher than their own (note that this eliminates delegation cycles). For each voter $i$ who decides to delegate, a `local delegation mechanism' takes in the competence levels of the voter's neighbours and returns a probability distribution over the delegations available to $i$. Delegations carry weight according to this probability distribution, and the collective decision is made by the majority rule. \citeauthor{Kang2018} prove a negative result: there is no local delegation mechanism which is strictly more accurate than allowing voters to vote directly.

\subsection{Structure of the Delegation Graph}

The models discussed in this section typically divide transitive proxy voting into three stages. In the first stage, a `delegation graph' (a graph representing delegations between voters) is formed from voters' preferences over alternatives and/or delegates. In the second stage, a preference profile is formed from this graph. In the third stage, this profile is used as an input to an aggregator. The model I propose in this paper works in the same way.

It is possible, though, to consider questions regarding the three stages independently. \textcite{Golz2018} focus on the first stage, the formation of the delegation graph from information about voters' delegation preferences. In the model they propose, each voter specifies a subset of the other voters who they would be happy to delegate their vote to. They consider the problem of assigning each voter a delegate from within this subset so as to minimise the number of voters delegating their vote to a single proxy (so as to minimise the maximum voting power of a voter who votes directly).

\textcite{Boldi} focus on the second stage; specifically, they consider the problem of forming a profile from the delegation graph in a way which minimises the maximum power an individual voter can accrue. Given a set of voters arranged in a network, their innovation is a `viscous' delegation factor $\alpha \in (0,1)$, representing the extent to which a delegation between neighbours in the network preserves voting power. The smaller $\alpha$ is, the more weight is lost every time a vote is transferred, so fine-tuning $\alpha$ could affect the feasible length of delegation chains. They discuss the impact of the structure of an underlying social network on the number of possible winners.

\section{Introducing the Model}
\label{introducing the model}
\subsection{Proxy Selection}

It's not my aim to give a full account of the factors that go into a voter's choice of proxy. That said, given the discussion in the previous section, I will take it that (1) and (2) are plausible starting points for an account of proxy selection:
\begin{itemize}
    \item[(1)] There is a large range of factors which inform a voter's choice of proxy (for example, a perception of competence, intelligence, honesty, etc). 
    \item[(2)] Voters pick proxies whom they think will represent their interests.
\end{itemize}

\paragraph{Example of Proxy Selection.} Suppose I am asked to give an ordinal ranking over the available options for the UK's future relationship with the European Union. I know that I prefer remaining in the EU to the other options, but I am unsure how to compare various intermediate levels of integration. If given the option of choosing a trusted delegate to submit an opinion on my behalf, I will do so.

I know that my friend Alice is exceptionally well informed about the intricacies of the EU. Ceteris paribus, then, she would be an excellent candidate to be my delegate (this is (1)).

I learn, though, that Alice prefers leaving the EU without a deal to remaining in the EU. The fact that Alice prefers a no-deal Brexit to remaining in the EU doesn't make me think that she's any less informed, or trustworthy, and so on, but it is sufficient to ensure that I won't pick her as my proxy. Since she disagrees with me so strongly on the issues on which I have made up my mind, I don't think she will represent my interests if she votes on my behalf (this is (2)).
\\
\\
\noindent I use this example to show that a model of proxy selection should have at least two interacting components. First and foremost, voters will only consider delegates who represent their interests (this is (2)). Beyond this, it is futile to attempt to place restrictions on their choice amongst potential delegates who they feel represent their interests, since many factors are relevant to this decision (this is (1)).

\subsection{Notation}

For a finite set $X$, let $\mathcal{P}(X)$ denote the set of all binary relations on $X$ which are irreflexive, anti-symmetric and transitive.

I will call $P\in \mathcal{P}(X)$ a `partial order', to emphasise that $P$ need not be total. Technically, of course, the relation usually called a `partial order' is reflexive rather than irreflexive. The reader should be mindful of this terminological idiosyncrasy, but it makes no difference to the content of this paper.

Following \textcite{Brill2018PairwiseLD}, I represent a partial order as a set of strict pairwise comparisons. This affects the notation I use. Suppose $X=\{a,b,c\}$. Then, using my terminology, the following are all examples of partial orders on $X$:
\begin{itemize}
    \item $P=\emptyset$
    \item $P'=\{a \succ b\}$
    \item $Q = \{a \succ b, a \succ c\}$
\end{itemize}
but the following would not be a partial order, since it is not closed under transitivity (since it doesn't contain $a \succ c$):
\begin{itemize}
    \item $Q' = \{a \succ b, b \succ c\}$
\end{itemize}
I will also speak of specific pairwise comparisons (or `edges') being members of a partial order. For example, I will say that $P'$ contains the edge $a\succ b$. Formally, I will write that $a\succ b \in P'$ (or, equivalently, that $\{a\succ b\}\subseteq P'$), but $a\succ b \notin P$ ($\{a\succ b\}\nsubseteq P$).

Let $\mathcal{L}(X)$ denote the set of all binary relations on $X$ which are irreflexive, anti-symmetric, transitive and also complete. Then I call $L\in \mathcal{L}(X)$ a `linear order'. Note that, by definition, $\mathcal{L}(X)\subseteq\mathcal{P}(X)$.

Throughout the paper, I will speak of profiles of partial (linear) orders. We can think of a profile of partial orders as a list of partial orders, one for each voter. So if $N=\{1,...,n\}$ is the set of voters, and $A$ is the set of alternatives, then $\boldsymbol{P}=(P_{1},...,P_{n})\in \mathcal{P}(A)^{n}$ is a list of partial orders (note I use the bold type face for lists of orders). By $P_{i}$, I designate the partial order submitted by voter $i$.

Fix some $\boldsymbol{P}=(P_{1},...,P_{n})$. Then, as is standard, we can also write $\boldsymbol{P}=(P_{i},\boldsymbol{P}_{-i})$ or $\boldsymbol{P}=(P_{i}, P_{j}, \boldsymbol{P}_{-i,j})$, for some $i,j\in N$. I will write $(P'_{i},\boldsymbol{P}_{-i})$ to designate the profile that is an `$i$-variant' of $(P_{i},\boldsymbol{P}_{-i})$. The same notational conventions apply to profiles of linear orders.

\subsection{Properties of Social Choice Functions}

I will use the phrase `social choice function' to refer to a resolute social choice function $f:\mathcal{L}(A)^{n}\ra A$ (so the reader can assume that the functions I consider have some sort of tie-breaking system built in). There are various familiar properties of social choice functions which will be relevant (there are many references for these standard axioms; see, for example, \cite{zwicker_moulin_2016}). Recall that $N$ is the set of voters $i,j,k, ...$ (with $|N|=n$), and $A$ is the set of alternatives $a,b,c, ...$ (with $|A|=m$).

\begin{definition}[Anonymity]
    A social choice function $f$ is \textit{anonymous} if, for any bijection $\psi:N\ra N$ and profile $\boldsymbol{L}=(L_{1},...,L_{n})\in \mathcal{L}(A)^{n}$, we have that
    \begin{equation*}
        f(L_{1},...,L_{n}) = f(L_{\psi(1)},...,L_{\psi(n)})
    \end{equation*}
\end{definition}

\noindent Let $\psi:A\ra A$ be a bijection. Let $P\in \mathcal{P}(A)$. By $\psi(P)$, I denote the alternative-wise application of the bijection. So if $P=\{a\succ b\}$, $\psi(a)=b$ and $\psi(b)=a$, then $\psi(P)=\{b\succ a\}$.

\begin{definition}[Neutrality]
    A social choice function $f$ is \textit{neutral} if, for any bijection $\psi:A\ra A$ and profile $\boldsymbol{L}=(L_{1},...,L_{n})\in \mathcal{L}(A)^{n}$, we have that
    \begin{equation*}
        \psi(f(L_{1},...,L_{n})) = f(\psi(L_{1}),...,\psi(L_{n}))
    \end{equation*}
\end{definition}

\begin{definition}[Weak Monotonicity]\label{Monotonicity}
    A social choice function $f$ is \textit{weakly monotonic} if the following holds for every $\boldsymbol{L}\in \mathcal{L}(A)^{n}$. Suppose $f(\boldsymbol{L})=a$, for some $a \in A$. Let $\boldsymbol{L'}=(L'_{i},\boldsymbol{L}_{-i})$ be an i-variant of $\boldsymbol{L}$, where
    \begin{equation*}
        L'_{i} = L_{i}\backslash \{b \succ a\} \cup \{a \succ b\}
    \end{equation*}
    for some $b \in A$ (in other words, voter $i$ moves alternative $a$ up at most one place in her ordering). Then we have that $f(\boldsymbol{L'})=a$.
\end{definition}

\noindent I will also define a novel property of social choice functions, which I will make use of in Section \ref{manipulation}.

Let $\boldsymbol{L+}$ be the profile we get when we augment $\boldsymbol{L}$ with $|A|!$ new voters, one holding each possible ranking in $\mathcal{L}(A)$ (if $f$ is not anonymous, we must assume some ordering on the rankings in $\mathcal{L}(A)$).

\begin{definition}[Uniform Voter Addition Invariance]
    A social choice function $f$ is \textit{Uniform Voter Addition Invariant} (UVAI) iff $f(\boldsymbol{L+})=f(\boldsymbol{L})$ for every $\boldsymbol{L}\in \mathcal{L}(A)^{n}$.
\end{definition}

\subsection{Proxy Votes}

My model extends a classical vote $(N,A,f)$ with a proxy mechanism, $g$. Recall that $\mathcal{P}(A)$ denotes the set of all partial orders over $A$. By $\mathscr{P}(N)$, I designate the powerset of $N$.

\begin{definition}[Proxy Mechanism]
    A function
    \begin{equation*}
        g: \mathcal{P}(A)^{n}\times N \rightarrow \mathscr{P}(N)
    \end{equation*}
    is a proxy mechanism iff, for every $\boldsymbol{P}=(P_{1},...,P_{n})\in \mathcal{P}(A)^{n}$, for every $i \in N$:
    \begin{itemize}
        \item[1.] If $P_{i}=\emptyset$, then $g(\boldsymbol{P},i)=N\backslash \{i\}$.
        \item[2.] If $P_{i}\in \mathcal{L}(A)$, then $g(\boldsymbol{P}, i)=\{i\}$.
        \item[3.] If $P_{i}\notin \mathcal{L}(A)$, then $i\notin g(\boldsymbol{P},i)$.
    \end{itemize}
\end{definition}
\noindent Intuitively, a proxy mechanism takes in a profile of partial orders and assigns to each voter a set of `permitted proxies', the voters whom they are allowed to choose as their delegate. The idea is that this set of permitted proxies constitutes the delegates who represent the voter's interests. So the proxy mechanism is designed to represent the features of proxy selection described above.

Firstly, if voter $i$ submits an empty order, we require (in 1.) that she can choose any other voter as her proxy (every other voter is in her set of permitted proxies). This is because she has no preferences over the alternatives, implying that there is no way for a potential delegate to fail to represent her interests.

Secondly, if voter $i$ submits a linear order, we require (in 2.) that she casts her own vote (she is the only voter in her set of permitted proxies). If she has already made her mind up about the alternatives, there is no need for her to delegate her vote to another voter.

Finally, if voter $i$ submits a partial order which is not a linear order, then she is not allowed (by 3.) to cast her own vote (she does not appear in her set of permitted proxies). This is because the social choice function takes profiles of linear orders as inputs; the model I propose modifies the method of collecting preferences, not the method of aggregating preferences.

So a proxy vote is a tuple $(N,A,f,g)$. Each voter $i \in N$ submits a triple $(P_{i},S_{i},D_{i})$, where:
\begin{itemize}
    \item $P_{i}\in \mathcal{P}(A)$ is a partial order over the alternatives. So the model allows voters to have made their mind up about some pairwise comparisons but not others.
    \item $S_{i}\in \mathcal{L}(N)$ is a linear order over the voters. Intuitively, this order corresponds to a ranking over potential proxies (capturing all the reasons that $i$ might have to prefer a delegate as her proxy independently of the delegate's ability to represent her views).
    \item $D_{i}\in \mathcal{L}(A)$ is a linear order over the set of alternatives, with $P_{i}\subseteq D_{i}$. $D_{i}$ is a `default vote'. In the situation where $i$ has no permitted proxies (so $g(\boldsymbol{P},i)=\emptyset$), $i$ is required to vote directly, submitting this default vote.
\end{itemize}

\noindent When each voter submits a triple, we have a \textit{proxy vote profile} $(\boldsymbol{P},\boldsymbol{S},\boldsymbol{D})$, where $\boldsymbol{P}$ is a (partial) \textit{preference profile}, $\boldsymbol{S}$ is a \textit{proxy choice profile} and $\boldsymbol{D}$ is a \textit{default vote profile}.

Each voter $i$ then receives $g(\boldsymbol{P},i)$, a set of permitted proxies, given the preference profile.

If $g(\boldsymbol{P},i)=\emptyset$, then $i$ must submit her default vote $D_{i} \in \mathcal{L}(A)$.

If $g(\boldsymbol{P},i)\neq\emptyset$, then $i$ must pick some $j \in g(\boldsymbol{P},i)$ to cast her vote on her behalf. Let $N'\subseteq N$. Then by $S_{i}|_{N'}$ I denote the restriction of $S_{i}$ to $N'$. Voter $i$ will pick the potential proxy who is ranked highest when we consider $S_{i}|_{g(\boldsymbol{P},i)}$ (in other words, the most preferred delegate from amongst her permitted proxies). Suppose that this is $j$. Then I will abuse notation by writing that $S_{i}|_{g(\boldsymbol{P},i)}=j$. For the sake of convenience, I will write $S_{i}|_{\{i\}}=i$ and $S_{i}|_{\emptyset}=i$, since $i$ casts her own vote if $g(\boldsymbol{P},i)=\{i\}$ or if $g(\boldsymbol{P},i)=\emptyset$.

So, given a voting profile $\boldsymbol{P}=(P_{1},...,P_{n})$ and proxy choice profile $\boldsymbol{S}=(S_{1},...,S_{n})$, each $i \in N$ has a proxy. So we have a \textit{delegation graph} $(N, R)$ where $iRj$ iff
\begin{equation*}
    j = S_{i}|_{g(\boldsymbol{P},i)}
\end{equation*}
Note that, where it does not have a negative impact on accuracy, I will speak of `$i$ choosing $j$ to be her proxy' as expressing this formal condition.
    
Let $R^{*}$ be the transitive closure of $R$. For each $i$, let
    \begin{equation*}
        \Pi_{i} = \{j \in N\:|\: iR^{*}j \text{ and } jRj\}
    \end{equation*}
    
If $\Pi_{i}$ is non-empty, it is easy to see that it will be a singleton $\{\pi_{i}\}$. Call $\pi_{i}$ voter $i$'s \textit{guru}. Note that if $i$ casts her own vote, then we have $\pi_{i}=i$ (so $i$ will be her own guru). We can then define a guru voting profile
    \begin{equation*}
        \boldsymbol{P_{\pi, \boldsymbol{S},\boldsymbol{D}}}=(P_{\pi_{1},\boldsymbol{S},\boldsymbol{D}},...,P_{\pi_{n},\boldsymbol{S},\boldsymbol{D}})
    \end{equation*}
Where $P_{\pi_{i},\boldsymbol{S},\boldsymbol{D}}$ is the preference order submitted by voter $i$'s guru $\pi_{i}$, generated according to $(\boldsymbol{P},\boldsymbol{S},\boldsymbol{D})$.

I use the notation $\boldsymbol{P_{\pi, \boldsymbol{S},\boldsymbol{D}}}$ to emphasise that this profile results from the proxy vote profile $(\boldsymbol{P},\boldsymbol{S},\boldsymbol{D})$. The use of $\boldsymbol{\pi}$ is supposed to remind the reader that the votes are actually submitted by the gurus $\pi_{1},...,\pi_{n}$. Note that, by construction, $P_{\pi_{i},\boldsymbol{S},\boldsymbol{D}}\in \mathcal{L}(A)$ for every $i \in N$, since each guru must cast her own vote. So we can use $\boldsymbol{P_{\pi, \boldsymbol{S},\boldsymbol{D}}}$ as the input to a social choice function. The \textit{outcome} of the proxy vote is given by $f(\boldsymbol{P_{\pi, \boldsymbol{S},\boldsymbol{D}}})$.\footnote{One might worry that there is a prohibitively large effort involved in submitting $S_{i}$ and $D_{i}$. To defend the cost of $D_{i}$, note that the default vote can be thought of as little more than a placeholder, a device which serves some practical purpose but has little ideological significance (e.g.\ each voter extends $P_{i}$ at random, or chooses the lexicographically earliest extension of $P_{i}$). To defend $S_{i}$, note that any real world version of transitive proxy voting will be situated within a dynamic environment. If we assume that the model is a static representation of a process which is inherently dynamic, then we might interpret the situation described by the model as follows. A voter $i$ submits $P_{i}$. She then calculates $g(\boldsymbol{P},i)$, the set of proxies she feels represent her interests, using the preferences submitted by the other voters. If $g(\boldsymbol{P},i)=\emptyset$, she submits $D_{i}\supset P_{i}$, her default preference. If $g(\boldsymbol{P},i)\neq\emptyset$, she picks some $j \in g(\boldsymbol{P},i)$ to be her proxy. So, rather than $S_{i}$, the voter $i$ is really only required to specify the name of a proxy in $g(\boldsymbol{P},i)$. It's true that the calculation of $g(\boldsymbol{P},i)$ will require some computation, but this shouldn't surprise us: the process of choosing a proxy does take some effort from the voter!}

\subsection{Examples of Proxy Mechanisms}

It is worth giving some examples of simple proxy mechanisms. By definition, for every proxy mechanism $g$ we have that $g(\boldsymbol{P},i)=N\backslash\{i\}$ if $P_{i}=\emptyset$, and $g(\boldsymbol{P},i)=\{i\}$ if $P_{i}\in \mathcal{L}(A)$. So it suffices to define the mechanism for the case where $P_{i}\neq \emptyset$ and $P_{i}\notin \mathcal{L}(A)$.

\begin{definition}(\texttt{TRIV})
\begin{equation*}
\texttt{TRIV}(\boldsymbol{P},i) = \emptyset \quad \text{ if } P_{i}\neq \emptyset \text{ and } P_{i}\notin \mathcal{L}(A)
\end{equation*}
If $g=$\texttt{ TRIV}, then every voter will cast her own vote, unless she has no preferences at all over the alternatives (in which case she will delegate). So we are close to a classical vote; proxy selection plays little role here.
\end{definition}

\begin{definition}(\texttt{UNIV})
\begin{equation*}
\texttt{UNIV}(\boldsymbol{P},i) = N\backslash\{i\} \quad \text{ if } P_{i}\neq \emptyset \text{ and } P_{i}\notin \mathcal{L}(A)
\end{equation*}
If $g=$\texttt{ UNIV}, then each voter who has not made her mind up fully can delegate her vote to any other voter, regardless of what she thinks on the issues she has made her mind up on. In effect, this is the formal set up of many of the models we discussed in the previous section; the strictly partial components of the preference profile $\boldsymbol{P}$ are irrelevant to proxy selection.
\end{definition}

\begin{definition}(\texttt{SUBSET})
\begin{equation*}
\texttt{SUBSET}(\boldsymbol{P},i) = \{j \in N\backslash\{i\}\:|\: P_{i}\subseteq P_{j}\} \quad \text{ if } P_{i}\neq \emptyset \text{ and } P_{i}\notin \mathcal{L}(A)
\end{equation*}
If $g=$\texttt{ SUBSET}, each voter who has not made up her mind fully can delegate to those voters whose preferences include her own as a subset.
\end{definition}

\begin{definition}(\texttt{DICTATOR})
For each $i\in N$, fix some $j_{i} \in N\backslash\{i\}$. Then
\begin{equation*}
\texttt{DICTATOR}(\boldsymbol{P},i) = \{j_{i}\} \quad \text{ if } P_{i}\neq \emptyset \text{ and } P_{i}\notin \mathcal{L}(A)
\end{equation*}
If $g=$\texttt{ DICTATOR}, then each voter $i$ has a unique dictator $j_{i}$; when $i$ submits some but not all pairwise comparisons, she must delegate her vote to $j_{i}$. \texttt{DICTATOR} won't be used in the remainder of the paper, but I define it to remind the reader that a proxy mechanism can act very differently for each voter it acts upon.
\end{definition}

\subsection{Representational Power}

I want to make three points about the representational power of the model I have presented.

Firstly, a proxy vote $(N,A,f,g)$ is a generalisation of a classical vote $(N,A,f)$. In the case where $P_{i}\in \mathcal{L}(A)$ for every $i \in N$, every voter casts her vote directly; we have a classical vote. In particular, this implies that any impossibility result concerning social choice functions $f$ will carry over into this setting. So, for example, the Gibbard-Sattherthwaite Theorem (\cite{Gibbard1973}, \cite{Satterthwaite1975}) holds in this novel setting.

Secondly, the model permits an easy resolution to delegation cycles. Following \textcite{Christoff17}, I have had voters submit a default vote which extends their existing vote. One could simply specify that this default vote is submitted directly by any voter who features in a delegation cycle. Thus delegation cycles require that the voters involved submit more pairwise comparisons in their votes, but do not prevent them from voting.

Thirdly, the model permits at least two ways of constraining delegations through arranging voters in a network (something explored by \textcite{Boldi}, \textcite{Golz2018} and \textcite{Bloembergen2018OnRD}). Fix some social network $(N,T\subseteq N\times N)$. Then we can build the social network into the range of the proxy mechanism $g$, requiring that $g(\boldsymbol{P},i)\subseteq T[i]$ for all $i\in N$, so that a voter can only delegate to her neighbours in the network. Another option is to place a requirement on $\boldsymbol{S}$, the proxy choice profile. For example, we could stipulate that
\begin{equation*}
    (j,k)\in S_{i} \text{ iff } (j\in T[i] \text{ and } k \notin T[i])
\end{equation*}
for every $i \in N$. What this says is that $i$ will always delegate to one of her neighbours if they are in her set of permitted proxies, but is able to delegate further afield if none of her neighbours represents her sufficiently.

\section{Properties of Proxy Mechanisms}
\label{properties of proxy mechanisms}
In this section, some natural properties of proxy mechanisms are defined. Some of these properties are then used to characterise the \texttt{SUBSET} mechanism.

\subsection{Defining Properties of Proxy Mechanisms}

Recall that a proxy mechanism is a function
\begin{equation*}
        g: \mathcal{P}(A)^{n}\times N \rightarrow \mathscr{P}(N).
\end{equation*}

\noindent Let $\psi:N\ra N$ be a bijection. For $N'\subseteq N$, I write $\psi(N')$ to denote the image of $N'$ under $\psi$. Let $\boldsymbol{P}\in \mathcal{P}(A)^{n}$ be a partial preference profile. Abusing notation, I write
\begin{equation*}
    \psi(\boldsymbol{P}) = \psi(P_{1},...,P_{n}) = (P_{\psi(1)},..., P_{\psi(n)})
\end{equation*}

\begin{definition}(Proxy Mechanism Anonymity)
    A proxy mechanism $g$ is \textit{anonymous} iff for every preference profile $\boldsymbol{P}\in \mathcal{P}(A)^{n}$ and every bijection $\psi:N\ra N$, we have that
    \begin{equation*}
        \psi(g(\boldsymbol{P},i)) = g(\psi(\boldsymbol{P}),\psi(i))
    \end{equation*}
\textit{Proxy Mechanism Anonymity} says that if we rename the voters, then a renamed voter's set of permitted proxies will just be the original voter's set of permitted proxies renamed. In other words, the proxy mechanism is blind to the identity of the individual voters.
\end{definition}

Let $\psi:A\ra A$ be a bijection. For $P\in \mathcal{P}(A)$, I write $\psi(P)$ to denote the alternative-wise application of the bijection. So if $P=\{a\succ b\}$, $\psi(a)=b$ and $\psi(b)=a$, then $\psi(P)=\{b\succ a\}$. Let $\boldsymbol{P}\in \mathcal{P}(A)^{n}$ be a partial preference profile. Abusing notation, I write $\psi(\boldsymbol{P}) = \psi(P_{1},...,P_{n}) = (\psi(P_{1}),...,\psi(P_{n}))$

\begin{definition}(Proxy Mechanism Neutrality)
    A proxy mechanism $g$ is \textit{neutral} iff for every preference profile $\boldsymbol{P}\in \mathcal{P}(A)^{n}$ and every bijection $\psi:A\ra A$, we have that
    \begin{equation*}
        g(\boldsymbol{P},i) = g(\psi(\boldsymbol{P}),i)
    \end{equation*}
\textit{Proxy Mechanism Neutrality} says that we can rename the alternatives without affecting each voter's set of permitted proxies.
\end{definition}

\begin{definition}(Proxy Availability (PA))
    $g$ satisfies \textit{PA} iff for every $P_{i}\in \mathcal{P}(A)$, for every $i \in N$, there is some $\boldsymbol{P}_{-i} \in \mathcal{P}(A)^{n-1}$ such that
    \begin{equation*}
        g((P_{i}, \boldsymbol{P}_{-i}),i)\neq \emptyset
    \end{equation*}
\textit{Proxy Availability (PA)} says that every voter should be able to find potential proxies for her votes, regardless of what views she holds, in at least some profile. In other words, every voter is capable of being represented, regardless of her views.
\end{definition}

\begin{definition}(Independence of Irrelevant Proxies (IIP))\label{IIP}
    $g$ satisfies \textit{IIP} iff for every $\boldsymbol{P},\boldsymbol{P'}\in \mathcal{P}(A)^{n}$, for every $i,j\in N$, if $P_{i}=P'_{i}$ and $P_{j}=P'_{j}$, then
    \begin{equation*}
        j \in (\boldsymbol{P},i) \text{ iff } j \in g(\boldsymbol{P'},i)
    \end{equation*}
\textit{Independence of Irrelevant Proxies (IIP)} says that whether $j$ is a permitted proxy for $i$ should depend only on $i$'s
and $j$'s preferences, not on the preferences of the other voters. It should be clear that IIP is motivated by the conception of proxy selection that I have argued for above.
\end{definition}

\begin{definition}(Zero Regret (ZR))\label{ZR}
    $g$ satisfies \textit{ZR} iff there is no triple $(\boldsymbol{P},\boldsymbol{S},\boldsymbol{D})$ (where $\boldsymbol{P}\in \mathcal{P}(A)^{n}$, $\boldsymbol{S}\in \mathcal{L}(N)^{n}$ and $\boldsymbol{D}\in \mathcal{L}(A)^{n}$) such that, for some $i \in N$:
    \begin{equation*}
        P_{i}\nsubseteq P_{\pi_{i},\boldsymbol{S},\boldsymbol{D}}
    \end{equation*}
\textit{Zero Regret (ZR)} says that a proxy mechanism guarantees that every voter's vote ends up being cast by someone who agrees with them completely (i.e.\ that they have no regrets about the vote submitted by their guru).
\end{definition}

\noindent Let $P_{i},Q_{i}\in \mathcal{P}(A)$. Then I write
        $Agree(P_{i},Q_{i}) = \{a\succ b \in P_{i} \:|\: a\succ b \in Q_{i}\}$
    and
        $Disagree(P_{i},Q_{i}) = \{a\succ b \in P_{i} \:|\: b\succ a \in Q_{i}\}$.

\begin{definition}(Preference Monotonicity (PM))\label{PM}
    $g$ satisfies \textit{PM} iff the following condition holds for every $\boldsymbol{P}\in \mathcal{P}(A)^{n}$, for every $i\in N$.
    Suppose $j\in g(\boldsymbol{P},i)$ and $j\neq i$. Then for every $k\in N\backslash\{i\}$, if
    \begin{equation*}
        Agree(P_{i},P_{j})\subseteq Agree(P_{i}, P_{k})
    \end{equation*}
    and
    \begin{equation*}
        Disagree(P_{i},P_{k})\subseteq Disagree(P_{i}, P_{j})
    \end{equation*}
    then $k \in g(\boldsymbol{P},i)$.
\end{definition}

\noindent \textit{Preference Monotonicity (PM)} says that if $j$ is a permitted proxy for $i$ and $k$ agrees with $i$ on at least the same things as $j$ whilst disagreeing with $i$ on at most the same things as $j$, then $k$ should also be a permitted proxy for $i$. It should be clear that PM is motivated by the account of proxy selection I have presented.

\subsection{Characterising \texttt{SUBSET}}

Recall that \texttt{SUBSET} was defined as follows:

\begin{equation*}
\texttt{SUBSET}(\boldsymbol{P},i) =
    \begin{cases}
        N\backslash\{i\} &\text{ if } P_{i}=\emptyset\\
        \{i\} &\text{ if } P_{i}\in \mathcal{L}(A)\\
        \{j \in N\backslash\{i\}\:|\: P_{i}\subseteq P_{j}\} &\text{ otherwise}
    \end{cases}
\end{equation*}

\begin{theorem}\label{characterising SUBSET}
\texttt{SUBSET} is the unique proxy mechanism satisfying \textit{Proxy Availability}, \textit{Independence of Irrelevant Proxies}, \textit{Zero Regret} and \textit{Preference Monotonicity}.
\end{theorem}

\begin{proof}
Clearly, \texttt{SUBSET} satisfies \textit{PA}, \textit{IIP} and \textit{ZR}. To see that \texttt{SUBSET} satisfies \textit{PM}, suppose that $j\in \texttt{SUBSET}(\boldsymbol{P},i)$ and $j\neq i$, for some $\boldsymbol{P} \in \mathcal{P}(A)^{n}$, $i,j \in N$. So $P_{i}\subseteq P_{j}$. Suppose that there is $k \in N$ such that $Agree(P_{i},P_{j})\subseteq Agree(P_{i},P_{k})$ and $Disagree(P_{i},P_{k})\subseteq Disagree(P_{i},P_{j})$. Then we must have $P_{i}\subseteq P_{k}$, since $Agree(P_{i},P_{j})=P_{i}$, since $P_{i}\subseteq P_{j}$. So $k \in \texttt{SUBSET}(\boldsymbol{P},i)$, as required.

For the other direction (i.e.\ to show uniqueness), I prove the contrapositive. Suppose $g\neq \texttt{SUBSET}$ is a proxy mechanism, and suppose $g$ satisfies \textit{PA}, \textit{IIP} and \textit{PM}. I will show that $g$ does not satisfy \textit{ZR}.

It will help to prove the following intermediate claim.

\begin{lemma}\label{One direction of SUBSET}
Let $\boldsymbol{P} \in \mathcal{P}(A)^{n}$ and $i,j \in N$, such that $P_{i}\notin \mathcal{L}(A)$. Then if $P_{i}\subseteq P_{j}$, and $g$ satisfies \textit{PA}, \textit{IIP} and \textit{PM}, we have $j \in g(\boldsymbol{P},i)$.
\end{lemma}
\noindent Since $g$ satisfies \textit{PA}, there must be some $\boldsymbol{P'}\in \mathcal{P}(A)^{n}$ such that $P'_{i}=P_{i}$ and $g(\boldsymbol{P'},i)\neq \emptyset$. Suppose $k \in g(\boldsymbol{P'},i)$, for some $k \in N$. Then we can construct a new profile $\boldsymbol{P''}$ where
\begin{align*}
    P''_{i} &= P'_{i}=P_{i}\\
    P''_{j} &= P_{j}\\
    P''_{k} &= P'_{k}
\end{align*}
By \textit{IIP}, we must have $k \in g(\boldsymbol{P''},i)$, since $P''_{i}=P'_{i}$ and $P''_{k}=P'_{k}$. But then by \textit{PM}, we must have $j \in (\boldsymbol{P''},i)$, since
\begin{equation*}
    P''_{i}=P_{i}\subseteq P_{j} = P''_{j}
\end{equation*}
implying that $j$ must agree at least as much with $i$ as $k$ in profile $\boldsymbol{P''}$. But then by another application of \textit{IIP}, we must have $j \in g(\boldsymbol{P},i)$, since $P''_{i}=P_{i}$ and $P''_{j}=P_{j}$. So Lemma \ref{One direction of SUBSET} holds.

We are now ready to prove the uniqueness of $\texttt{SUBSET}$. Since $g\neq \texttt{SUBSET}$, there must be some $\boldsymbol{P}\in \mathcal{P}(A)^{n}$ and $i,j\in N$ with $P_{i}\notin \mathcal{L}(A)$ such that either
\begin{equation*}
    P_{i}\nsubseteq P_{j} \text{ and } j \in g(\boldsymbol{P},i)
\end{equation*}
or
\begin{equation*}
    P_{i}\subseteq P_{j} \text{ and } j \notin g(\boldsymbol{P},i)
\end{equation*}
But note that Lemma \ref{One direction of SUBSET} rules out this latter case. So we only need to consider the case where $P_{i}\nsubseteq P_{j} \text{ and } j \in g(\boldsymbol{P},i)$.
Since $P_{i}\nsubseteq P_{j}$, there must be some $a\succ b\in P_{i}$ such that $a\succ b\notin P_{j}$.

But now consider a profile $\boldsymbol{P'}$ where, for some $k \in N$:
\begin{align*}
    P'_{i}&=P_{i}\\
    P'_{j}&=P_{j}\\
    P'_{j}\cup \{b\succ a\}&\subseteq P'_{k}, \text{ and } P'_{k}\in \mathcal{L}(A)
\end{align*}
Note that this profile is well defined; since $a\succ b\notin P_{j}=P'_{j}$, we must have that $P'_{j}\cup \{b\succ a\}$ is still anti-symmetric, and thus can be extended to a linear order $P'_{k}$.

Since $P'_{i}=P_{i}$ and $P'_{j}=P_{j}$, we must have $j\in g(\boldsymbol{P'},i)$, by \textit{IIP}. We must also have $k\in g(\boldsymbol{P'},j)$ by Lemma \ref{One direction of SUBSET}, since $P'_{j}\subseteq P'_{k}$. So then if $i$ picks $j$ as her proxy and $j$ picks $k$ as her proxy, then $k$ will be $i$'s guru. But $P'_{i}\nsubseteq P'_{k}$, since $a\succ b\in P_{i}=P'_{i}$ and $b\succ a\in P'_{k}$. So $g$ is not \textit{ZR}. 
\end{proof}

\noindent In fact, \textit{PA}, \textit{IIP}, \textit{ZR} and \textit{PM} are all individually necessary for characterising \texttt{SUBSET}.

\texttt{TRIV} (defined in Section \ref{introducing the model}) is a proxy mechanism which does not satisfy \textit{PA}. But note that \texttt{TRIV} does satisfy \textit{IIP}, \textit{ZR} and \textit{PM}.

\texttt{UNIV} (defined in Section \ref{introducing the model}) is a proxy mechanism which does not satisfy \textit{ZR}. But note that \texttt{UNIV} does satisfy \textit{PA}, \textit{IIP} and \textit{PM}.

Consider $g$ defined as follows:
\begin{equation*}
g(\boldsymbol{P},i) =
    \begin{cases}
    N\backslash\{i\} &\text{ if } P_{i}=\emptyset\\
        \{i\} &\text{ if } P_{i}\in \mathcal{L}(A)\\
        \{j \in N\:|\: P_{j}\in \mathcal{L}(A) \text{ and } P_{i}\subset P_{j}\}  &\text{ otherwise}
    \end{cases}
\end{equation*}
$g$ is a proxy mechanism which does not satisfy \textit{PM} (just consider some $P_{j}\notin \mathcal{L}(A)$ such that $P_{i}\subseteq P_{j}$). But note that $g$ does satisfy \textit{PA}, \textit{IIP} and \textit{ZR}.

Consider $g$ defined as follows:
\begin{equation*}
g(\boldsymbol{P},i) =
    \begin{cases}
    N\backslash\{i\} &\text{if } P_{i}=\emptyset\\
        \{i\} &\text{if } P_{i}\in \mathcal{L}(A)\\
        \{j \in N\:|\: P_{i}\subseteq P_{j}\}  &\text{if } P_{i}\notin \mathcal{L}(A) \textit{ and } (\forall j\in N\backslash\{i\} \text{ such that }\\ &P_{j} \in \mathcal{L}(A)) \text{ we have } P_{i}\subseteq P_{j} \\
        \emptyset &\text{otherwise }
    \end{cases}
\end{equation*}
$g$ is a proxy mechanism which does not satisfy \textit{IIP}. But note that $g$ does satisfy \textit{PA}, \textit{PM} and \textit{ZR}.

\section{Properties of Proxy Votes}
\label{properties of proxy votes}
In this section, properties of pairs $(f,g)$ are examined.

\subsection{Defining Properties of Proxy Votes}

Suppose $\psi:N\ra N$ is a bijection. Suppose $\boldsymbol{P}\in \mathcal{P}(A)^{n}$ is a preference profile. Then it will be convenient to write
\begin{equation*}
    \psi(\boldsymbol{P})=(P_{\psi(1)},...,P_{\psi(n)})
\end{equation*}
to denote the voter-wise application of the bijection $\psi$. Likewise with a default vote profile $\boldsymbol{D}\in \mathcal{L}(A)^{n}$.

Suppose $S\in \mathcal{L}(N)$. Then I write $\psi(S)$ to denote the voter-wise permutation of $S$. For example, if $S=\{i\succ j\}$, then $\psi(S)=\{\psi(i)\succ \psi(j)\}$.

Suppose $\boldsymbol{S}\in \mathcal{L}(N)^{n}$ is a proxy choice profile. Then, abusing notation, I write
\begin{equation*}
    \psi(\boldsymbol{S})=(\psi(S)_{\psi(1)},...,\psi(S)_{\psi(n)})
\end{equation*}
to denote the voter-wise application of the bijection $\psi$ to both the profile and the content of each voter's proxy choice.

\begin{definition}(Proxy Vote Anonymity)
    A pair $(f,g)$, where $f$ is a social choice function and $g$ is a proxy mechanism, satisfies \textit{Proxy Vote Anonymity} iff for every $\boldsymbol{P}\in \mathcal{P}(A)^{n}$, for every $\boldsymbol{S}\in \mathcal{L}(N)^{n}$ for every $\boldsymbol{D}\in \mathcal{L}(A)^{n}$, and for every bijection $\psi:N\ra N$:
    \begin{equation*}
        f(\boldsymbol{P_{\pi,S,D}}) = f(\psi(\boldsymbol{P})_{\boldsymbol{\pi},\psi(\boldsymbol{S}),\psi(\boldsymbol{D})})
    \end{equation*}
\textit{Proxy Vote Anonymity} says that renaming the voters does not affect the result of the proxy vote.
\end{definition}

Suppose $\psi:A\ra A$ is a bijection. Let $P\in\mathcal{P}(A)$. Then I write $\psi(P)$ to denote the alternative-wise permutation of $P$. For example, if $P=\{a\succ b\}$, then $\psi(P)=\{\psi(a)\succ \psi(b)\}$.

Suppose $\boldsymbol{P}\in \mathcal{P}(A)^{n}$ is a (partial) preference profile. Then, abusing notation, I write
\begin{equation*}
    \psi(\boldsymbol{P})=(\psi(P_{1}),...,\psi(P_{n}))
\end{equation*}
to denote the voter-wise application of the bijection $\psi$. Likewise with a default vote profile $\boldsymbol{D}\in \mathcal{L}(A)^{n}$.

\begin{definition}(Proxy Vote Neutrality)
    A pair $(f,g)$, where $f$ is a social choice function and $g$ is a proxy mechanism, satisfies \textit{Proxy Vote Neutrality} iff for every $\boldsymbol{P}\in \mathcal{P}(A)^{n}$, for every $\boldsymbol{S}\in \mathcal{L}(N)^{n}$, for every $\boldsymbol{D}\in \mathcal{L}(A)^{n}$ and for every bijection $\psi:A\ra A$:
    \begin{equation*}
        \psi(f(\boldsymbol{P_{\pi,S,D}})) = f(\psi(\boldsymbol{P})_{\boldsymbol{\pi},\boldsymbol{S},\psi(\boldsymbol{D})})
    \end{equation*}
\textit{Proxy Vote Neutrality} says that renaming the alternatives just renames the outcome of the proxy vote.
\end{definition}

\noindent In a proxy vote setting, voters submit partial orders over alternatives. This means there are two ways they can increase their support for an alternative $a\in A$. They can either add an edge $a \succ b$, or remove an edge $b \succ a$. With this in mind, we can distinguish between two notions of monotonicity in a proxy vote setting: `addition monotonicity' and `deletion monotonicity'. As an anonymous reviewer observes, the proxy vote analogue of the classical monotonicity property can be analysed as a conjunction of these two notions.

\begin{definition}(Proxy Vote Addition Monotonicity (PVAM))\label{Proxy Vote Addition Monotonicity}
    A pair $(f,g)$, where $f$ is a social choice function and $g$ is a proxy mechanism, satisfies \textit{PVAM} iff the following holds for every $\boldsymbol{P}\in \mathcal{P}(A)^{n}$, every $\boldsymbol{S}\in \mathcal{L}(N)^{n}$ and every $\boldsymbol{D}\in \mathcal{L}(A)^{n}$.
    
    Suppose $f(\boldsymbol{P_{\pi,S,D}})=a$ for some $a \in A$. Consider an i-variant of $\boldsymbol{P}$, $\boldsymbol{P'}=(P'_{i},\boldsymbol{P}_{-i})$, where $P'_{i}=P_{i}\cup \{a\succ b\}$, for some $b \in A$. Then $f(\boldsymbol{P'_{\pi,S,D}})=a$.
\end{definition}

\noindent \textit{Proxy Vote Addition Monotonicity (PVAM)} says that if the winner under some proxy vote profile $(\boldsymbol{P},\boldsymbol{S},\boldsymbol{D})$ is $a$, and we modify $\boldsymbol{P}$ by having some voter add a pairwise comparison to favour $a$, then the winner should remain $a$.

\begin{definition}(Proxy Vote Deletion Monotonicity (PVDM))\label{Proxy Vote Deletion Monotonicity}
    A pair $(f,g)$, where $f$ is a social choice function and $g$ is a proxy mechanism, satisfies \textit{PVDM} iff the following holds for every $\boldsymbol{P}\in \mathcal{P}(A)^{n}$, every $\boldsymbol{S}\in \mathcal{L}(N)^{n}$ and every $\boldsymbol{D}\in \mathcal{L}(A)^{n}$.
    
    Suppose $f(\boldsymbol{P_{\pi,S,D}})=a$, for some $a \in A$. Consider an i-variant of $\boldsymbol{P}$, $\boldsymbol{P'}=(P'_{i},\boldsymbol{P}_{-i})$, where $P'_{i}=P_{i}\backslash \{b\succ a\}$, for some $b \in A$. Then $f(\boldsymbol{P'_{\pi,S,D}})=a$.
\end{definition}

\noindent \textit{Proxy Vote Deletion Monotonicity (PVDM)} says that if the winner under some proxy vote profile $(\boldsymbol{P},\boldsymbol{S},\boldsymbol{D})$ is $a$, and we modify $\boldsymbol{P}$ by having some voter delete a pairwise comparison which favours some other alternative over $a$, then the winner should remain $a$.

\subsection{Interaction with Properties of $f$ and $g$}

Having defined properties of pairs $(f,g)$, it's interesting to explore how they relate to properties of the individual components $f$ and $g$. The first two results follow immediately from the relevant definitions; simply consider preference profiles where voters submit linear preferences.

\begin{proposition}\label{Proxy vote anonymity implies anonymity}
If $(f,g)$ satisfies \textit{proxy vote anonymity}, then $f$ satisfies \\\textit{anonymity}.
\end{proposition}

\begin{proposition}\label{Proxy vote neutrality implies neutrality}
If $(f,g)$ satisfies \textit{proxy vote neutrality}, then $f$ satisfies \textit{neutrality}.
\end{proposition}

\begin{proposition}\label{Proxy vote addition and deletion monotonicity implies monotonicity}
If $(f,g)$ satisfies \textit{proxy vote addition monotonicity} and \textit{proxy vote deletion monotonicity}, then $f$ satisfies \textit{weak monotonicity}.
\end{proposition}

\begin{proof}
By contraposition. Suppose $f$ is not weakly monotonic. We will show that either $(f,g)$ fails to satisfy PVAM or $(f,g)$ fails to satisfy PVDM.

Since $f$ is not weakly monotonic, there must be $\boldsymbol{P},\boldsymbol{P'}\in \mathcal{L}(A)^{n}$, where $\boldsymbol{P'}=(P'_{i},\boldsymbol{P}_{-i})$ and
    \begin{equation*}
        P'_{i} = P_{i}\backslash\{b \succ a\} \cup \{a \succ b\}
    \end{equation*}
such that $f(\boldsymbol{P})=a$ and $f(\boldsymbol{P'})\neq a$, for some $a,b\in A$.

Define $P''_{i}=P_{i} \backslash \{b\succ a\}$. By definition, $P'_{i}=P''_{i}\cup \{a\succ b\}$. Define $\boldsymbol{P''}=(P''_{i},\boldsymbol{P}_{-i})$. Fix some arbitrary proxy choice profile $\boldsymbol{S}$ and default vote profile $\boldsymbol{D}$. Then $\boldsymbol{P}_{\boldsymbol{\pi},\boldsymbol{S},\boldsymbol{D}}=\boldsymbol{P}$ and $\boldsymbol{P'}_{\boldsymbol{\pi},\boldsymbol{S},\boldsymbol{D}}=\boldsymbol{P'}$, by construction. So $f(\boldsymbol{P}_{\boldsymbol{\pi},\boldsymbol{S},\boldsymbol{D}})=a$ and $f(\boldsymbol{P'}_{\boldsymbol{\pi},\boldsymbol{S},\boldsymbol{D}})\neq a$. If $f(\boldsymbol{P''}_{\boldsymbol{\pi},\boldsymbol{S},\boldsymbol{D}})=a$, then $(f,g)$ fails to satisfy PVAM (since adding the edge $a\succ b$ changes the winner from $a)$. If $f(\boldsymbol{P''}_{\boldsymbol{\pi},\boldsymbol{S},\boldsymbol{D}})\neq a$, then $(f,g)$ fails to satisfy PVDM (since removing the edge $b\succ a$ has changed the winner from $a$). 
\end{proof}

\begin{lemma}\label{Anonymous + Proxy Mechanism Anonymous implies Proxy Vote Anonymous}
If $f$ is anonymous and $g$ is proxy mechanism anonymous, then $(f,g)$ is proxy vote anonymous.
\end{lemma}

\begin{proof}
Let a proxy vote profile $(\boldsymbol{P},\boldsymbol{S},\boldsymbol{D})$ be arbitrary. Pick some bijection $\psi:N\ra N$. Then we must have
\begin{align*}
    f(\psi(\boldsymbol{P})_{\boldsymbol{\pi},\psi(\boldsymbol{S}),\psi(\boldsymbol{D})}) &= f(\psi(\boldsymbol{P}_{\boldsymbol{\pi},\boldsymbol{S},\boldsymbol{D}})) &\text{(since $g$ is anonymous)} & \\
    &= f(\boldsymbol{P}_{\boldsymbol{\pi},\boldsymbol{S},\boldsymbol{D}}) &\text{(since $f$ is anonymous)} &\quad 
\end{align*}
\end{proof}

\begin{lemma}\label{Neutral + Proxy Mechanism Neutral implies Proxy Vote Neutral}
If $f$ is neutral and $g$ is proxy mechanism neutral, then $(f,g)$ is proxy vote neutral.
\end{lemma}

\begin{proof}
Let a proxy vote profile $(\boldsymbol{P},\boldsymbol{S},\boldsymbol{D})$ be arbitrary. Pick some bijection $\psi:A\ra A$. Then we must have
\begin{align*}
    f(\psi(\boldsymbol{P})_{\boldsymbol{\pi},\boldsymbol{S},\psi(\boldsymbol{D})}) &= f(\psi(\boldsymbol{P}_{\boldsymbol{\pi},\boldsymbol{S},\boldsymbol{D}})) &\text{(since $g$ is neutral)} & \\
    &= \psi(f(\boldsymbol{P}_{\boldsymbol{\pi},\boldsymbol{S},\boldsymbol{D}})) &\text{(since $f$ is neutral)} &\quad 
\end{align*}
\end{proof}

\noindent Note that, in general, the other direction of Lemmas \ref{Anonymous + Proxy Mechanism Anonymous implies Proxy Vote Anonymous} and \ref{Neutral + Proxy Mechanism Neutral implies Proxy Vote Neutral} won't hold.

\subsection{A Proxy Vote Analogue of May's Theorem}

When $|A|=2$, the \textit{majority rule} selects the alternative which receives the most first choice votes. When $|N|$ is odd (meaning the majority rule is resolute), May (\cite{May1952ASO}) shows that we can characterise the majority rule as the unique rule satisfying anonymity, neutrality and weak monotonicity.

We can use the proxy vote analogues of these properties to achieve the same characterisation result. The key point is that setting $|A|=2$ fully specifies the proxy mechanism $g$, by the definition of a proxy mechanism (since either a voter submits an empty order, meaning she is allowed to delegate to any other voter, or she submits a linear order, meaning she casts her vote directly). In effect, we are close to a classical vote; the only voters who delegate their votes are the voters who submit empty orders. So it is unsurprising that the following result holds irrespective of the choice of $g$.

\begin{theorem}\label{May's Theorem}
Suppose $|A|=2$ and $|N|$ is odd. Then a pair $(f,g)$ satisfies
\begin{itemize}
    \item Proxy Vote Anonymity
    \item Proxy Vote Neutrality
    \item Proxy Vote Addition Monotonicity (PVAM), and
    \item Proxy Vote Deletion Monotonicity (PVDM)
\end{itemize}
iff $f$ is the majority rule.
\end{theorem}

\begin{proof}
The left to right direction follows from Propositions \ref{Proxy vote anonymity implies anonymity}, \ref{Proxy vote neutrality implies neutrality}, \ref{Proxy vote addition and deletion monotonicity implies monotonicity} and May's Theorem.

For the other direction, suppose that $f$ is the majority rule. So $f$ is anonymous and neutral. When $|A|=2$, note there is only a single proxy mechanism $g$, by the definition of a proxy mechanism. So $g$ will be anonymous and neutral. By Lemmas \ref{Anonymous + Proxy Mechanism Anonymous implies Proxy Vote Anonymous} and \ref{Neutral + Proxy Mechanism Neutral implies Proxy Vote Neutral}, this implies that $(f,g)$ is anonymous and neutral.

It remains only to show that $(f,g)$ satisfies PVAM and PVDM. I will write $A=\{a,b\}$. Suppose that for some proxy vote profile $(\boldsymbol{P},\boldsymbol{S},\boldsymbol{D})$, we have
\begin{equation*}
    f(\boldsymbol{P}_{\boldsymbol{\pi},\boldsymbol{S},\boldsymbol{D}}) =a
\end{equation*}
Fix some $i \in N$. Then there are two cases to consider.

To see that $(f,g)$ satisfies PVAM, suppose that $P_{i}=\emptyset$, then consider the case where $P'_{i}=\{a\succ b\}$. So $i$ casts her own vote, meaning $P'_{\pi_{i},\boldsymbol{S},\boldsymbol{D}}=\{a\succ b\}$. Note that if $j$ picked $i$ as her proxy when $P_{i}=\emptyset$, then $j$ must still pick $i$ as her proxy (since this implies $P_{j}=\emptyset$, since $|A|=2$). And we know that $P_{\pi_{i},\boldsymbol{S},\boldsymbol{D}}$ was either $\{b\succ a\}$ or $\{a\succ b\}$.

To see that $(f,g)$ satisfies PVDM, suppose that $P_{i}=\{b\succ a\}$, then consider the case where $P'_{i}=\emptyset$. So $i$ delegates her vote, meaning $P'_{\pi_{i},\boldsymbol{S},\boldsymbol{D}}$ is either $\{a\succ b\}$ or $\{b \succ a\}$. Note that if $j$ didn't pick $i$ as her proxy when $P_{i}=\{b\succ a\}$, then $j$ must still not pick $i$ as her proxy (since this implies either that $P_{j}\in \mathcal{L}(A)$, or that $P_{j}=\emptyset$ and $j$ prefers some other voter in $N\backslash\{i\}$). And we know that $P_{\pi_{i},\boldsymbol{S},\boldsymbol{D}}$ was $\{b\succ a\}$, since $i$ cast her own vote.

In either case, changing from $P_{i}$ to $P'_{i}$ can only decrease the number of $\{b\succ a\}$ edges submitted in the profile $(P'_{i},\boldsymbol{P}_{-i})_{\boldsymbol{\pi},\boldsymbol{S},\boldsymbol{D}}$ from in the profile $\boldsymbol{P}_{\boldsymbol{\pi},\boldsymbol{S},\boldsymbol{D}}$ and increase the number of $\{a\succ b\}$ edges submitted in the profile $(P'_{i},\boldsymbol{P}_{-i})_{\boldsymbol{\pi},\boldsymbol{S},\boldsymbol{D}}$ from in the profile $\boldsymbol{P}_{\boldsymbol{\pi},\boldsymbol{S},\boldsymbol{D}}$. Since $f$ is weakly monotonic, this implies that $f((P'_{i},\boldsymbol{P}_{-i})_{\boldsymbol{\pi},\boldsymbol{S},\boldsymbol{D}})=a$. So $(f,g)$ satisies PVAM and PVDM. 
\end{proof}

\subsection{Proxy Vote Monotonicity: An Impossibility Result}

Given some plausible restrictions on $g$, the monotonicity property of $g$ turns out to be incompatible with monotonicity properties of the pair $(f,g)$ for a large class of social choice functions.

A social choice function $f$ is a \textit{scoring rule} if it can be expressed as a vector $(s_{1},...,s_{m})$ with $s_{1}\geq ... \geq s_{m} \geq 0$ and $s_{1}>s_{m}$. Each $a\in A$ receives $s_{p}$ points for each voter putting it in the $p$th position in her ballot, and the outcome is the alternative $a$ with the most points $s_{a}$.

\begin{theorem}\label{Big impossibility result}
Suppose $|A|=3$. Then, for $|N|\geq14$, there is no pair $(f,g)$, where $f$ is a scoring rule and $g$ is a proxy mechanism, such that:
\begin{itemize}
    \item $(f,g)$ satisfies \textit{proxy vote addition monotonicity (PVAM)} and \textit{proxy vote deletion monotonicity (PVDM)}
    \item $g$ satisfies \textit{preference monotonicity (PM)} and \textit{independence of irrelevant proxies (IIP)}
\end{itemize}
\end{theorem}

I write $|A|=\{a,b,c\}$. Let $f$ be a scoring rule, and assume $g$ satisfies PM and IIP. I show that $(f,g)$ must either fail to satisfy PVAM or fail to satisfy PVDM.

The following lemma is at the core of the proof. It says that for any proxy mechanism satisfying PM and IIP there must exist profiles where adding an edge $\{a \succ b\}$ or removing an edge $\{b \succ a \}$ switches some voter $i$'s guru's vote from $\{b \succ a \succ c\}$ to $\{c \succ a \succ b\}$ (for some $a,b,c\in A$). I will show that this implies that at least one of the monotonicity properties fails for scoring rules, completing the proof of Theorem \ref{Big impossibility result}. Note, though, that Lemma \ref{big impossibility result lemma} is a result about proxy mechanisms $g$; it is entirely independent of social choice functions $f$.

\begin{lemma}\label{big impossibility result lemma}
Let $|N|\geq 3$ and $|A|=3$. If $g$ satisfies PM and IIP, then, for some $i\in N$, for some $a,b,c\in A$, there exist $P_{i},P'_{i}\in \mathcal{P}(A)$, $\boldsymbol{P}_{-i}\in \mathcal{P}(A)^{n-1}$, $\boldsymbol{S}\in \mathcal{L}(N)^{n}$, $\boldsymbol{D}\in \mathcal{L}(A)^{n}$ such that:
\begin{itemize}
    \item Either $P'_{i}=P_{i}\cup \{a\succ b\}$ or $P'_{i}=P_{i}\backslash \{b \succ a\}$
    \item $(P_{i},\boldsymbol{P}_{-i})_{\pi_{i},\boldsymbol{S},\boldsymbol{D}} = \{b\succ a \succ c\}$
    \item $(P'_{i},\boldsymbol{P}_{-i})_{\pi_{i},\boldsymbol{S},\boldsymbol{D}} = \{c\succ a \succ b\}$
\end{itemize}
In other words, we can construct a profile where the vote cast by $i$'s guru changes from $\{b\succ a \succ c\}$ to $\{c\succ a \succ b\}$ when $i$'s vote changes from $P_{i}$ to $P'_{i}$.
\end{lemma}

\begin{proof}
\noindent There are three collectively exhaustive cases:\medskip

\hrule

\medskip

\noindent\textbf{Case 1:} For some $i\in N$, for some $a,b\in A$, for some $\boldsymbol{P}_{-i}\in \mathcal{P}(A)^{n-1}$: $P'_{i} = \{a \succ b\}$ and $g((P'_{i},\boldsymbol{P}_{-i}),i) \neq N\backslash\{i\}$.
\\

\noindent\textbf{Proof of Case 1:} Let $P_{i}=\emptyset$. So $P'_{i}=P_{i}\cup\{a \succ b\}$. We construct a profile where the vote cast by $i$'s guru is $\{b\succ a \succ c\}$ when she submits $P_{i}$ and $\{c\succ a \succ b\}$ when she submits $P'_{i}$.

By assumption, $g((P'_{i},\boldsymbol{P}_{-i}),i) \neq N\backslash\{i\}$. If $g((P'_{i},\boldsymbol{P}_{-i}),i)=\emptyset$, then let $D_{i}=\{c\succ a\succ b\}$ (in this case, $i$ would be her own guru when she casts the vote $P'_{i}$; so her guru would vote for $\{c\succ a\succ b\}$, as required). If $g((P'_{i},\boldsymbol{P}_{-i}),i)\neq \emptyset$, then pick some $j \in g((P'_{i},\boldsymbol{P}_{-i}),i)$. Let $P'_{j}=\{b\succ a \succ c\}$, and let $S_{i}|_{N\backslash\{i\}}=j$. Since $g$ satisfies IIP and $g((P'_{i},\boldsymbol{P}_{-i}),i) \neq N\backslash\{i\}$, we must have $g((P'_{i},P'_{j}, \boldsymbol{P}_{-i,j}),i) \neq N\backslash\{i\}$. To see this, consider some voter in $N\backslash\{j\}$ who was not a permitted proxy for $i$ before $j$ changed her vote; since neither this voter nor $i$ have changed their votes, it follows from the fact that $g$ is IIP that this voter must not be a permitted proxy for $i$ after $j$ changes her vote. Since $g$ satisfies PM, this implies that we must have $j \notin g((P'_{i},P'_{j}, \boldsymbol{P}_{-i,j}),i)$. To see this, note that $P'_{j}$ has the minimum number of agreements and maximum number of disagreements possible with $P'_{i}$. So if $j$ were a permitted proxy for $i$, then every other voter would have to be -- regardless of what she voted -- since $g$ is PM, contradicting our earlier reasoning.

Pick some $k \in N\backslash\{i,j\}$ and set $P'_{k}=\{c \succ a \succ b\}$. Since $g$ satisfies IIP, we must have $j \notin g((P'_{i},P'_{j},P'_{k}, \boldsymbol{P}_{-i,j,k}),i)$. If $g((P'_{i},P'_{j},P'_{k}, \boldsymbol{P}_{-i,j,k}),i)=\emptyset$, set $D_{i}=\{c\succ a\succ b\}$ (as above, $i$ would here be her own guru; so her guru would vote for $\{c\succ a\succ b\}$, as required). If $g((P'_{i},P'_{j},P'_{k}, \boldsymbol{P}_{-i,j,k}),i)\neq \emptyset$, then we must have $k \in g((P'_{i},P'_{j},P'_{k}, \boldsymbol{P}_{-i,j,k}),i)$, since $g$ satisfies PM. Set $S_{i}|_{g((P'_{i},P'_{j},P'_{k}, \boldsymbol{P}_{-i,j,k}),i)}=k$. So $i$ chooses $k$ as her proxy when she votes for $P'_{i}$.

Note that, since $P_{i}=\emptyset$, we have $g((P_{i},P'_{j},P'_{k}, \boldsymbol{P}_{-i,j,k}),i)=N\backslash\{i\}$ by definition. It follows that we must have $(P_{i},P'_{j},P'_{k}, \boldsymbol{P}_{-i,j,k})_{\pi_{i},\boldsymbol{S},\boldsymbol{D}} = \{b\succ a \succ c\}$, since $i$ will pick $j$ as her proxy in this case (since we have specified that $S_{i}|_{N\backslash\{i\}} = j$). By construction, we must have $(P'_{i},P'_{j},P'_{k}, \boldsymbol{P}_{-i,j,k})_{\pi_{i},\boldsymbol{S},\boldsymbol{D}} = \{c\succ a \succ b\}$, since $i$ will either pick $k$ as her proxy in this case or submit her default vote.

\medskip

\hrule

\medskip

\noindent \textbf{Case 2:} For every $i \in N$, $a,b\in A$, $\boldsymbol{Q}_{-i}\in \mathcal{P}(A)^{n-1}$, if $P_{i}=\{a\succ b\}$, we have $g((P_{i},\boldsymbol{Q}_{-i}), i) = N\backslash\{i\}$ (i.e. Case 1 doesn't hold). For some $i\in N$, for some $a,b,c\in A$, for some $\boldsymbol{P}_{-i}\in \mathcal{P}(A)^{n-1}$: $P'_{i} = \{a \succ b, c \succ b\}$ and $g((P'_{i},\boldsymbol{P}_{-i}),i) \neq N\backslash\{i\}$.
\\

\noindent\textbf{Proof of Case 2:} Let $P_{i}=\{c \succ b\}$. So $P'_{i}=P_{i}\cup \{a\succ b\}$. By assumption, $g((P'_{i},\boldsymbol{P}_{-i}),i)\neq N\backslash\{i\}$. If $g((P'_{i},\boldsymbol{P}_{-i}),i)=\emptyset$, then let $D_{i}=\{c\succ a\succ b\}$ (as in the proof of the previous case, $i$ is her own guru in this situation). Otherwise, let $P'_{j}=\{b \succ a \succ c\}$ for some $j \in g((P'_{i},\boldsymbol{P}_{-i}),i)$ and $P'_{k}=\{c \succ a \succ b\}$ for some $k \in N\backslash\{i,j\}$. Set $S_{i}|_{N\backslash\{i\}}=j$.

If $g((P'_{i},P'_{j},P'_{k}, \boldsymbol{P}_{-i,j,k}),i)$ $= \emptyset$, set $D_{i}=\{c\succ a\succ b\}$ (as above, $i$ is her own guru in this situation). Otherwise, set $S_{i}|_{g((P'_{i},P'_{j},P'_{k}, \boldsymbol{P}_{-i,j,k}),i)}=k$. Since $g$ satisfies IIP and PM, identical reasoning to that in Case 1 shows that $j \notin g((P'_{i},P'_{j},P'_{k}, \boldsymbol{P}_{-i,j,k}),i)$ and $k \in$ $g((P'_{i},P'_{j},P'_{k},$ $\boldsymbol{P}_{-i,j,k}),i)$.

Note that $g((P_{i},P'_{j},P'_{k}, \boldsymbol{P}_{-i,j,k}),i)=N\backslash\{i\}$, by the assumption that Case 1 is false. It follows that we must have $(P_{i},P'_{j},P'_{k}, \boldsymbol{P}_{-i,j,k})_{\pi_{i},\boldsymbol{S},\boldsymbol{D}} = \{b\succ a \succ c\}$, since $i$ will pick $j$ as her proxy in this case. By construction, we must have $(P'_{i},P'_{j},P'_{k}, \boldsymbol{P}_{-i,j,k})_{\pi_{i},\boldsymbol{S},\boldsymbol{D}}$ $= \{c\succ a \succ b\}$, since $i$ will either pick $k$ as her proxy in this case or submit her default vote.

\medskip

\hrule

\medskip

\noindent \textbf{Case 3:} For every $i \in N$, $a,b\in A$, $\boldsymbol{Q}_{-i}\in \mathcal{P}(A)^{n-1}$, if $P_{i}=\{a\succ b\}$, we have $g((P_{i},\boldsymbol{Q}_{-i}), i) = N\backslash\{i\}$ (i.e. Case 1 doesn't hold). For every $i \in N$, $a,b,c\in A$, $\boldsymbol{Q}_{-i}\in \mathcal{P}(A)^{n-1}$, if $P_{i}=\{a\succ c, b \succ c\}$, we have $g((P_{i},\boldsymbol{Q}_{-i}), i) = N\backslash\{i\}$ (i.e. Case 2 doesn't hold).
\\

\noindent\textbf{Proof of Case 3:} Let $P_{i}=\{b \succ a \succ c\}$ and $P'_{i}=\{a \succ c, b\succ c\}$. Then $P'_{i}=P_{i}\backslash\{b\succ a\}$. Let $P_{k}=\{c\succ a \succ b\}$ for some $k \in N\backslash\{i\}$. Let $S_{i}|_{N\backslash\{i\}}=k$, and fix some $\boldsymbol{P}_{-i,k}\in \mathcal{P}(A)^{n-2}$.

By construction, $(P_{i},P_{k}, \boldsymbol{P}_{-i,k})_{\pi_{i},S_{i},D_{i}} = \{b\succ a \succ c\}$, since $i$ casts her own vote (by the definition of a proxy mechanism, since $P_{i}$ is a linear order). By the assumption that Case 2 doesn't hold, $g((P'_{i},P_{k}, \boldsymbol{P}_{-i,k}),i)=N\backslash\{i\}$. So $k \in g((P'_{i},P_{k}, \boldsymbol{P}_{-i,k}),i)$. So $(P'_{i},P_{k}, \boldsymbol{P}_{-i,k})_{\pi_{i},\boldsymbol{S},\boldsymbol{D}} = \{c\succ a \succ b\}$, since $i$ delegates her vote to $k$ in this situation.

\medskip

\hrule

\medskip

\noindent Since the three cases are collectively exhaustive, Lemma \ref{big impossibility result lemma} holds.

What we have shown, then, is that there must exist profiles where adding an edge $\{a \succ b\}$ or removing an edge $\{b \succ a \}$ switches a voter $i$'s guru's vote from $\{b \succ a \succ c\}$ to $\{c \succ a \succ b\}$. In these profiles, we require that $P_{j}=\{b\succ a \succ c\}$ and $P_{k}=\{c\succ a \succ b\}$ for some $j,k\in N\backslash\{i\}$. Crucially, though, since $g$ is IIP, we are free to vary the votes of the voters in $N\backslash\{i,j,k\}$ as we wish whilst ensuring that $i$'s final vote will still change in the constructed way.

To complete the proof, all we need do is set the votes submitted by the voters in the set $N\backslash\{i,j,k\}$ to construct profiles where $a$ wins when $i$'s final vote is $\{b\succ a\succ c\}$, and $c$ wins when $i$'s final vote is $\{c\succ a \succ b\}$. We do this as follows.

Note that since we are dealing with resolute scoring rules, we require slightly different solutions depending on whether the tie break contains $a \succ c$ or $c \succ a$.\footnote{It should be clear that we could modify the proof to accommodate irresolute rules.}

For a tie break containing $a \succ c$, the following solution works for even $n$. Have two voters vote for $b \succ c \succ a$, one voter vote for $a \succ c \succ b$ and the remaining $n-6$ voters divide their vote evenly between $c \succ a \succ b$ and $a \succ c \succ b$.

If $i$'s guru votes for $b \succ a \succ c$, then:
\begin{align*}
    s_{a} &= (\frac{n}{2}-2)s_{1} + (\frac{n}{2})s_{2} + 2s_{3}\\
    s_{b} &= 4s_{1} + (n - 4)s_{3} \\
    s_{c} &= (\frac{n}{2}-2)s_{1} + (\frac{n}{2})s_{2} + 2s_{3}
\end{align*}
For $|N|\geq 14$, we must have that $s_{a},s_{c}> s_{b}$, since $s_{1}>s_{3}$ and $s_{2}\geq s_{3}$ by definition. So $a$ wins the election.

If $i$'s guru votes for $c \succ a \succ b$, then:
\begin{align*}
    s_{a} &= (\frac{n}{2}-2)s_{1} + (\frac{n}{2})s_{2} + 2s_{3}\\
    s_{b} &= 3s_{1} + (n - 3)s_{3} \\
    s_{c} &= (\frac{n}{2}-1)s_{1} + (\frac{n}{2})s_{2} + s_{3}
\end{align*}
So $s_{c}>s_{a}$, meaning $c$ wins the election.

For a tie break containing $a \succ c$ and odd $n$, the exact solution depends on the scoring rule. If $s_{1}>s_{2}=s_{3}$, then add one voter with $b\succ a \succ c$ to the solution for even $n$. If $s_{1}>s_{2}>s_{3}$ or $s_{1}=s_{2}>s_{3}$, add one voter with $a \succ c \succ b$ to the solution for even $n$. This exhausts the possible scoring rules.

For a tie break containing $c \succ a$, the following solution works for odd $n$. Have two voters vote for $b \succ c \succ a$, one voter vote for $a \succ c \succ b$, one voter vote for $a \succ b \succ c$ and the remaining $n-7$ voters divide their vote evenly between $c \succ a \succ b$ and $a \succ c \succ b$. It is easily verified that $s_{a}>s_{c}$ when $i$'s guru votes for $b \succ a \succ c$ (meaning $a$ wins), and $s_{a}=s_{c}$ when $i$'s guru votes for $c \succ a \succ b$ (meaning $c$ wins).

For a tie break containing $c \succ a$ and even $n$, the exact solution depends on the scoring rule. If $s_{1}>s_{2}=s_{3}$ or $s_{1}>s_{2}> s_{3}$, add one voter with $b \succ c \succ a$ to the solution for odd $n$. If $s_{1}=s_{2}>s_{3}$, add one voter with $a \succ c \succ b$ to the solution for odd $n$. This exhausts the possible scoring rules.

Any tie break must contain either $a\succ c$ or $c\succ a$. It follows that $(f,g)$ must either fail to satisfy PVAM or fail to satisfy PVDM. 

\end{proof}

\section{Manipulation}
\label{manipulation}
I turn now to the topic of manipulation. Manipulation has hitherto received little attention in the literature on liquid democracy. \citeauthor{Blum2016} (\citeyear{Blum2016}) mention manipulation as a potential issue with democratic systems, but the notion of manipulation they have in mind is that by the agenda-setter (often called `control' in the social choice literature), rather than by voters themselves. \citeauthor{Brill-InteractiveDemocracy} (\citeyear{Brill-InteractiveDemocracy}) notes that introducing delegation permits voters to manipulate the outcomes of votes in novel ways, but does not develop a formal model to support this claim. The most developed formal model of manipulation in a proxy vote setting comes from \citeauthor{escoffier_convergence_iterative} (\citeyear{escoffier_convergence_iterative}), discussed in Section \ref{related work}. Recall that in their model, voters submit preferences over potential gurus, but do not also submit preferences over some set of alternatives (that is, there is no background election against which the delegation takes place). The authors consider ways in which voters might misrepresent their preferences so as to obtain more preferred gurus. So the notion of manipulation they discuss is not manipulation of the outcome of an election, but rather of the endpoints of individual delegation chains. In particular, this implies that -- unlike the notions of manipulation defined in this section -- there is no relationship between their notion of manipulation and standard notions of manipulation.

In this section, a novel form of manipulation (`proxy choice manipulation') is defined, which is shown to occur roughly as often as classical manipulation. Classical manipulation is then generalised to the proxy vote setting (`preference misrepresentation manipulation'), and it is shown that manipulation occurs strictly more often in proxy votes.

\subsection{Proxy Choice Manipulation}

In a classical vote $(N,A,f)$, voters can manipulate by misrepresenting their preferences to achieve a better outcome. In a proxy vote $(N,A,f,g)$, there is an additional option for manipulation. Voters can manipulate by misrepresenting their choice of proxy (i.e.\ by picking one proxy over another for strategic reasons). I call this sort of manipulation `proxy choice manipulation'. Note that in a proxy vote setting, manipulability is no longer a property of a social choice function $f$ alone, but rather of a pair $(f,g)$.

\begin{definition}[Proxy Choice Manipulation]
    A pair $(f,g)$ is \textit{proxy choice manipulable} (PC-manipulable) iff there exists $i \in N$, $\boldsymbol{P}\in \mathcal{P}(A)^{n}$, $\boldsymbol{S}\in \mathcal{L}(N)^{n}$, $\boldsymbol{D}\in \mathcal{L}(A)^{n}$ such that:
    \begin{equation*}
        f(\boldsymbol{P}_{\pi,\boldsymbol{S},\boldsymbol{D}}) \prec f(\boldsymbol{P}_{\pi,(S'_{i},\boldsymbol{S}_{-i}),\boldsymbol{D}})\in P_{i}
    \end{equation*}
    for some $S_{i},S'_{i}\in\mathcal{L}(N)$.
\end{definition}

\noindent Intuitively, a pair $(f,g)$ is PC-manipulable if there is a profile where a voter would prefer one of her potential proxies over another for purely strategic reasons.

A natural question to investigate is how PC-manipulability relates to the standard notion of manipulability, which I'll call `Gibbard-Satterthwaite Manipulability' (GS-manipulability).

\begin{definition}[Gibbard-Satterthwaite Manipulation]
    A social choice function $f$ is \textit{Gibbard-Satterthwaite manipulable} (GS-manipulable) iff there exists $i \in N$, $\boldsymbol{P}_{-i}\in \mathcal{L}(A)^{n-1}$ such that:
    \begin{equation*}
        f((P_{i},\boldsymbol{P}_{-i}))\prec f((P'_{i},\boldsymbol{P}_{-i}))\in P_{i}
    \end{equation*}
    for some $P_{i},P'_{i}\in \mathcal{L}(A)$.
\end{definition}

\noindent One way of investigating the connection between PC-manipulability and GS-manipulability is to fix a particular proxy mechanism $g$.

\begin{theorem}\label{PC-manipulability implies GS-manipulability}
If $(f,\texttt{SUBSET})$ is PC-manipulable for $n$ voters and $m$ alternatives, then $f$ is GS-manipulable for $n$ voters and $m$ alternatives.
\end{theorem}

\begin{proof}
Suppose $(f,\texttt{SUBSET})$ is PC-manipulable for $n$ voters and $m$ alternatives. Then there is some preference profile $\boldsymbol{P}$, default profile $\boldsymbol{D}$ and proxy choices of the voters in $N\backslash\{i\}$, $\boldsymbol{S}_{-i}$, where $i$ strictly prefers the outcome of the vote when she submits $(P_{i}, S'_{i}, D_{i})$ to the outcome when she submits $(P_{i}, S_{i}, D_{i})$. Let $\boldsymbol{S}=(S_{i},\boldsymbol{S}_{-i})$ and $\boldsymbol{S'}=(S'_{i},\boldsymbol{S}_{-i})$.

Let
\begin{equation*}
    Proxy_{i} = \{j \in N\:|\: P_{\pi_{j},\boldsymbol{S},\boldsymbol{D}} \neq P_{\pi_{j},\boldsymbol{S'},\boldsymbol{D}}\}
\end{equation*}
be the set of voters whose guru's vote changes when $i$ changes her choice of proxy (i.e. the set of voters whose vote `flows through' $i$; note that $i \in Proxy_{i}$, by assumption).

Without loss of generality, suppose $f(\boldsymbol{P}_{\pi,\boldsymbol{S},\boldsymbol{D}}) = b$
and $f(\boldsymbol{P}_{\pi,\boldsymbol{S'},\boldsymbol{D}})=a$.
So $a\succ b\in P_{i}$. Since we are using the \texttt{SUBSET} mechanism, this implies that $a\succ b \in P_{\pi_{j},\boldsymbol{S},\boldsymbol{D}}$ for every $j\in Proxy_{i}$.

Suppose now that we move from the profile $\boldsymbol{P}_{\pi,\boldsymbol{S},\boldsymbol{D}}$ towards the profile $\boldsymbol{P}_{\pi,\boldsymbol{S'},\boldsymbol{D}}$ by changing, for each $j \in Proxy_{i}$, $P_{\pi_{j},\boldsymbol{S},\boldsymbol{D}}$ to $P_{\pi_{j},\boldsymbol{S'},\boldsymbol{D}}$.

We know that when we start, the outcome is $b$. We know that when we have made all the changes, the outcome is $a$. If the outcome changes directly from $b$ to $a$ at some stage in the process, then we have a profile with respect to which $f$ is GS-manipulable (since $a\succ b \in P_{\pi_{j},\boldsymbol{S},\boldsymbol{D}}$ for every $j\in Proxy_{i}$). If the social outcome first changes to some $c\neq b \neq a$, then there are two cases. If $c\succ b \in P_{\pi_{i},\boldsymbol{S},\boldsymbol{D}}$, then the same reasoning shows that $f$ is GS-manipulable. If $c\prec b \in P_{\pi_{i},\boldsymbol{S},\boldsymbol{D}}$, then we can just carry on making the changes until the social outcome changes to $a$, then apply the same reasoning as above. It follows that $f$ is GS-manipulable. 
\end{proof}

\noindent So we have shown that PC-manipulability implies GS-manipulability, assuming $g$ is the \texttt{SUBSET} mechanism. In general, the converse won't hold (just consider the case where $|N|=2$). But we can say something about the converse relationship, using a stronger form of GS-manipulability.

\begin{definition}[IIA-Manipulation]
    A social choice function $f$ is \textit{IIA- }\textit{manipulable} if there is some $\boldsymbol{L}\in \mathcal{L}(A)$ such that, for some $i\in N$, $L'_{i}\neq L_{i}$:
    \begin{itemize}
        \item $f(L'_{i},\boldsymbol{L}_{-i}) \succ f(\boldsymbol{L})\in L_{i}$
        \item $f(L'_{i},\boldsymbol{L}_{-i}) \succ f(\boldsymbol{L})\in L'_{i}$
    \end{itemize}
\end{definition}
\noindent Intuitively, $f$ is IIA-manipulable if a voter can reverse the social ranking of two alternatives whilst maintaining her personal ranking of the alternatives.\footnote{IIA-manipulability can be thought of as a much weaker condition than `one-way monotonicity' (\cite{Sanver_Zwicker_One_Way_Monotonicity}), which features in the preference reversal paradox (\cite{Peters17}). In effect, one-way monotonicity says that every example of GS-manipulability is an example of IIA-manipulability.}

\begin{theorem}\label{IIA-manipulability implies PC-manipulability}
If $f$ is:
\begin{itemize}
    \item IIA-manipulable over $n$ voters and $m$ alternatives.
    \item Invariant to Uniform Voter Additions
\end{itemize}
then $(f,\texttt{SUBSET})$ is PC-manipulable over $n+m!$ voters, and $m$ alternatives.
\end{theorem}

\begin{proof}
Suppose $f$ is IIA-manipulable over $|N|=n$ voters and $|A|=m$ alternatives. Note that we must have $m>2$, by the definition of IIA-manipulability. Since $f$ is IIA-manipulable, there must be some profile $\boldsymbol{L}\in \mathcal{L}(A)^{n}$ and some $i \in N$ such that, for some $L'_{i}\neq L_{i}$, we have
\begin{equation*}
    f(L'_{i},\boldsymbol{L}_{-i})\succ_{L_{i}} f(\boldsymbol{L})
\end{equation*}
For the sake of readability, let $f(\boldsymbol{L})=b$ and $f(L'_{i},\boldsymbol{L}_{-i})=a$. Since $f$ is IIA-manipulable, we can assume that both $a \succ b \in L_{i}$ and $a \succ b \in L'_{i}$ without loss of generality.

Let us now consider $\boldsymbol{L+}$ and $\boldsymbol{L'+}$, the uniform-voter augmentations of $\boldsymbol{L}$ and $(L'_{i},L_{-i})$ respectively. Since $f$ is IUVA, it follows that $f(\boldsymbol{L+}) = b$ and $f(\boldsymbol{L'+}) = a$. Since $\boldsymbol{L+}$ contains, for every linear order over $A$, at least one voter who submits that order, we must have voters $j$ and $k$ such that $L+_{j}=L_{i}$ and $L+_{k}=L'_{i}$.

Define $P_{i} = a \succ b$. Note that both $P_{i}\subset L+_{j}$ and $P_{i}\subset L+_{k}$. Since we are using the \texttt{SUBSET} proxy mechanism, this implies that both $j$ and $k$ are permitted proxies for $i$ in the preference profile $(P_{i},\boldsymbol{L+}_{-i})$.

If $i$ picks $j$ as her proxy, then the guru profile for $(P_{i},\boldsymbol{L+}_{-i})$, written as $(P_{i},\boldsymbol{L+}_{-i})_{\boldsymbol{\pi},\boldsymbol{S},\boldsymbol{D}}$, is simply $\boldsymbol{L+}$. So we must have $f((P_{i},\boldsymbol{L+}_{-i})_{\boldsymbol{\pi},\boldsymbol{S},\boldsymbol{D}})=b$.

If $i$ picks $k$ as her proxy, then the guru profile for $(P_{i},\boldsymbol{L+}_{-i})$, written as $(P_{i},\boldsymbol{L+}_{-i})_{\boldsymbol{\pi},\boldsymbol{S'},\boldsymbol{D}}$, is simply $\boldsymbol{L'+}$. So we must have $f((P_{i},\boldsymbol{L+}_{-i})_{\boldsymbol{\pi},\boldsymbol{S'},\boldsymbol{D}})=a$.

Since $a\succ b \in P_{i}$, it follows that we have a situation where $i$ would strictly prefer picking $k$ over $j$ as her proxy. So $f$ is PC-manipulable on a profile of $n+m!$ voters. 
\end{proof}

\subsection{Preference Misrepresentation Manipulation}

It is also natural to generalise GS-manipulation in the proxy vote setting.

\begin{definition}[Preference Misrepresentation Manipulation]
    A pair $(f,g)$ is \textit{preference misrepresentation manipulable} (PM-manipulable) iff there exists $i \in N$, $\boldsymbol{P}\in \mathcal{P}(A)^{n}$, $\boldsymbol{S}\in \mathcal{L}(N)^{n}$ such that:
    \begin{equation*}
        f(\boldsymbol{P}_{\pi,\boldsymbol{S},\boldsymbol{D}}) \prec f((P'_{i},\boldsymbol{P}_{-i})_{\pi,\boldsymbol{S},\boldsymbol{D}})\in P_{i}
    \end{equation*}
    for some $P_{i},P'_{i}\in\mathcal{P}(A)$.
\end{definition}

\noindent Given PM-manipulability is just the generalisation of GS-manipulability to the proxy vote setting, one might wonder whether domain restrictions which result in GS-strategyproofness also result in PM-strategyproofness. The following result shows that this does not hold. There are social choice functions $f$ such that $(f,\texttt{SUBSET})$ is PM-manipulable on the domain of single-peaked preference profiles but $f$ is not GS-manipulable on the domain of single-peaked preference profiles.

\begin{theorem}\label{PM-manipulability happens in single-peaked domains}
When $|A|= 3$ and $|N|\geq 2$, there is no social choice function $f$ which is non-dictatorial, surjective, and such that $(f,\textup{\texttt{SUBSET}})$ is PM-strategyproof, even when we restrict the domain to include only single-peaked preference profiles.
\end{theorem}

\begin{proof}
We know that if a social choice function $f$ is GS-manipulable, the pair $(f,\texttt{SUBSET})$ is PM-manipulable. Moulin characterises the class of surjective, non-dictatorial and GS-strategyproof social choice functions on the domain of single-peaked preferences as the class of generalised median voter rules (\cite{Black1948}, \cite{Moulin1980}). To prove the theorem at hand, then, it suffices to show that for a generalised median voter rule $f$, $(f,\texttt{SUBSET})$ is PM-manipulable on the domain of single-peaked preferences when $|A|= 3$. I write $A=\{a,b,c\}$.

Let $f$ be an arbitrary generalised median voter rule with $n-1$ phantoms. Without loss of generality, suppose that at least one phantom has peak $a$. We construct the profile $\boldsymbol{P}=(P_{1},...,P_{n})$, where
\begin{align*}
    P_{j} &= \{c \succ b\} &\quad \forall j \in N\backslash\{i\}\\
    P_{i} &= \{b \succ c \succ a\}
\end{align*}
Suppose also that $D_{j}=\{a \succ c \succ b\}$ for every $j \in N\backslash\{i\}$, and that $S_{j}|_{N\backslash\{j\}}=i$ (i.e. that every $j$ would pick $i$ as her proxy if permitted). Note that $\boldsymbol{P}$ is single-peaked along the dimension $a,c,b$.

As it stands, we have that $g(\boldsymbol{P},j)=N\backslash \{i,j\}$ for every $j \in N\backslash\{i\}$ (since $g$ is the \texttt{SUBSET} mechanism). It follows that each of these voters will enter a delegation cycle, casting her default vote $\{a \succ c \succ b\}$. So the peak of the median voter will be $a$, since $n - 1$ voters and at least one phantom have peak $a$. So the winner will be $a$.

Now suppose $i$ switches from $\{b\succ c\succ a\}$ to $\{c\succ a \succ b\}$. Note that the profile is still single-peaked along the dimension $a,c,b$. Now we have that $g(\boldsymbol{P},j)=N\backslash\{j\}$ for every $j \in N\backslash\{i\}$. By construction, it follows that each of the voters has $i$ as her guru. So the peak of all $n$ voters will be $c$, implying that the peak of the median voter will be $c$ (even if none of the phantoms has peak $c$). So the winner will be $c$. Since $c\succ a \in P_{i}$, it follows that $i$ has an incentive to change her preference from $\{b\succ c\succ a\}$ to $\{c\succ a \succ b\}$. 
\end{proof}

\section{Conclusion}
\label{conclusion}
This paper introduced a novel model of transitive proxy voting, which paid more attention to `proxy selection', the process by which voters select delegates. The properties of the model were explored from an axiomatic perspective; it was shown that (given plausible assumptions) we cannot expect proxy votes to satisfy intuitively desirable monotonicity properties. The model was also put to work in analysing manipulation in a proxy vote setting. It was shown not only that novel forms of manipulation arise in a proxy vote setting, but also that there are strictly more situations in which manipulation is available to voters in a proxy vote than in a classical vote.

A natural question concerns the relation between the main results in this paper and the informal arguments surrounding transitive proxy voting of the sort discussed in Section \ref{introduction}. Do these formal results have implications for real world discussion of liquid democracy?

Underpinning the model I've introduced is the claim that voters will only delegate their votes to proxies who represent their interests. I introduced the claim normatively (akin to a constraint on voters' rationality); in order for the claim to be testable empirically, we need to operationalise the notion of a proxy representing a voter's interests. Proxy mechanisms provide us with useful tools with which to do this -- different formal properties of proxy mechanisms will lend themselves to developing and evaluating competing notions of what it means for a proxy to represent a voter's interests. In particular, Theorem \ref{characterising SUBSET} (which characterises the \texttt{SUBSET} mechanism) can be seen as providing a formal argument for the plausible empirical claim that a voter will only delegate to a proxy who agrees with her on all the issues on which she's already made up her mind.

When there are only two alternatives, it's natural to think that the degree to which a proxy vote resembles a classical vote depends entirely on the number of voters who choose to delegate rather than vote directly. Since it supports this intuition, Theorem \ref{May's Theorem} (the proxy vote analogue of May's Theorem) is best viewed as justification for the formal model introduced, rather than as a result with practical implications of its own.

Given plausible assumptions on the process by which voters select proxies, Theorem \ref{Big impossibility result} shows that proxy votes will always fail to satisfy monotonicity properties for a large class of monotonic social choice functions. In effect, it serves to highlight the instability inherent to votes involving delegations; small changes in an individual voter's behaviour can change the outcomes of votes in counterintuitive ways. My sense is that Theorem \ref{Big impossibility result} ought to be taken seriously by opponents of liquid democratic systems. Recall that a key motivation for transitive proxy voting is that it is more `democratic', in that it better represents the views of the whole electorate (\cite{Green-Armytage2015}). But a failure of monotonicity is precisely a situation in which an aggregation procedure has represented the views of an electorate poorly. Similarly, it is often claimed that proxy voting increases participation (\cite{Miller1969}, \cite{Alger2006}). One might think that the sort of reasoning employed in the proof of Theorem \ref{Big impossibility result} suggests that the model proposed in this paper is well positioned to challenge this participation claim formally (future work could do just this).

As noted in Section \ref{manipulation}, it has been claimed informally that transitive proxy voting equips voters with the ability to manipulate in ways unavailable to them in classical votes (\cite{Blum2016}, \cite{Brill-InteractiveDemocracy}). The model proposed in this paper confirms such claims, and provides a formal framework with which to examine the material consequences of novel forms of manipulation. Some of the results in this section will reassure proponents of liquid democracy; for example, Theorems \ref{PC-manipulability implies GS-manipulability} and \ref{IIA-manipulability implies PC-manipulability} challenge the idea that a voter's novel ability to misrepresent her choice of proxy gives her any additional power to manipulate the outcome of the election. Another result, though, lends support to arguments advanced against liquid democracy; Theorem \ref{PM-manipulability happens in single-peaked domains} shows that outcomes of proxy votes are strictly more vulnerable than those of classical votes to manipulation by an individual voter who misrepresents her preferences. Furthermore, the proof of Theorem \ref{PM-manipulability happens in single-peaked domains} exploits the fact that voting power in a transitive proxy voting system can concentrate in the hands of individual `super-voters', a common worry raised against liquid democratic systems (\cite{Kling2015}). Future work could apply the model in this paper to a wider range of manipulation and control problems.

My aim in this paper has not been to provide full formal coverage of topics relevant to transitive proxy voting, but rather to showcase interesting features of the model I've introduced. My hope is that the reader thinks my model sufficiently rich to enable non-trivial formal discussion of the arguments surrounding transitive proxy voting.

\begin{acknowledgements}
I'd like especially to thank Ulle Endriss, who provided whole hosts of fruitful suggestions at every stage in the writing of this paper. Thanks also to Davide Grossi and Ronald de Haan for generous feedback on an earlier draft.
\end{acknowledgements}

\printbibliography
\end{document}